\documentclass[journal]{IEEEtran}
\pdfoutput=1
\usepackage{blindtext}
\usepackage{graphicx}
\usepackage{float}
\usepackage[english]{babel}
\usepackage{color}
\usepackage{fp}
\usepackage{calc}
\usepackage{amsmath}
\usepackage{array}
\usepackage{cite}
\usepackage{xcolor}
\usepackage{rotating}
\usepackage[percent]{overpic}
\usepackage{turnstile}
\usepackage{tikz}
\usepackage{enumitem}
\usepackage{lipsum}
\usepackage{soul}
\usetikzlibrary{shapes,arrows}
\usepackage{siunitx}

\usepackage[switch, columnwise]{lineno}

\definecolor{darkgreen}{RGB}{10,100,20}
\begin{document}
\bibliographystyle{IEEEtranS}
\raggedbottom
\title{{M}odeling ultrasound propagation in the moving brain: applications to shear shock waves and traumatic brain injury}
\author{Sandhya~Chandrasekaran,
        Bharat~B.~Tripathi,
         David~Esp\'indola,
         and~Gianmarco~F.~Pinton%
\thanks{B.B. Tripathi, D. Esp\'indola and G. Pinton are with the Joint Department
of Biomedical Engineering, University of North Carolina at Chapel Hill and the North Carolina State University, Chapel Hill, North Carolina. \newline D. Esp\'indola is currently with Instituto de Ciencias de la Ingenier\'ia, Universidad de O'Higgins, Avenida Libertador Bernardo O'Higgins 611, Rancagua, Chile. \newline S. Chandrasekaran is with the Department of Mechanical and Aerospace Engineering, North Carolina State University, North Carolina.}
}


\maketitle
\begin{abstract}
\textcolor{black}{Traumatic Brain Injury (TBI) studies on the living human brain are experimentally infeasible due to ethical reasons and the elastic properties of the brain degrade rapidly post-mortem.} We present a simulation approach that models ultrasound propagation in the human brain while it is moving due to the complex shear shock wave deformation from a traumatic impact. Finite difference simulations can model ultrasound propagation in complex media such as human tissue. Recently, we have shown that the Fullwave finite difference approach can also be used to represent displacements that are much smaller than the grid size, such as the motion encountered in shear wave propagation from ultrasound elastography. However, this sub-resolution displacement model, called impedance flow, was only implemented and validated for acoustical media composed of randomly distributed scatterers. Here we propose a generalization of the impedance flow method that describes the continuous sub-resolution motion of structured acoustical maps, and in particular of acoustical maps of the human brain. It is shown that the average error in the subresolution displacement method is small compared to the wavelength {($\lambda/1702$)}. The method is then applied to acoustical maps of the human brain with a motion that is imposed by the propagation of a shear shock wave. This motion is determined numerically with a custom piecewise parabolic method that is calibrated to {\it ex vivo} observations of shear shocks in the porcine brain. Then the Fullwave simulation tool is used to model transmit-receive imaging sequences based on an {L7-4} imaging transducer. The simulated radiofrequency data is beamformed using a conventional delay-and-sum method and a normalized cross-correlation method designed for shock wave tracking is used to determine the tissue motion.  {This overall process is an  {\it in silico} reproduction of the experiments that were previously performed to observe shear shock waves in fresh porcine brain.} It is shown that the proposed generalized impedance flow method accurately captures the shear wave motion in terms of the wave profile, shock front characteristics, odd harmonic spectrum generation, and acceleration at the shear shock front. We expect that this approach will lead to improvements in image sequence design that takes into account the aberration and multiple reflections from the brain and in the design of tracking algorithms that can more accurately capture the complex brain motion that occurs during a traumatic impact. These methods of modeling ultrasound propagation in moving media can also be applied to other displacements, such as those generated by shear wave elastography or blood flow.




\end{abstract}

\IEEEpeerreviewmaketitle

\section{Introduction}




\IEEEPARstart{T}raumatic Brain Injury (TBI)  {studies must often rely on simulations since experiments  that generate injury in the human brain are infeasible. Furthermore, experiments that rely on the elastic properties of the brain are challenging due to the rapid degradation of neural tissue post-mortem. We present a simulation approach to generate ultrasound images of the brain undergoing nonlinear elastic motion due to a traumatic impact.} 
Finite difference simulations can model ultrasound propagation through the multi-scale heterogeous representation of human tissue \cite{Mast1997,mast1998effect,Mast2002,pinton2009heterogeneous}.  Finite Difference Time Domain (FDTD) methods can directly model wave propagation and hence, account for physical phenomena such as  single and multiple scattering, reflection and refraction.  The backscattered energy received at the transducer surface can then be beamformed into a highly realistic ultrasound image. Using time domain equations, simulations of linear and lossless propagation of ultrasound in maps of the human abdominal wall have been successful in visualizing the time-shift aberration in the wavefront due to large-scale tissue inhomogeneities \cite{mast1998effect}.
 {A FDTD-based simulation tool (referred herein as ``Fullwave'') that we have previously developed \cite{pinton2009heterogeneous}, uses a second-order method to describe nonlinear propagation of acoustical waves in heterogeneous, attenuating  media, such as the soft tissue of the human body.} The Fullwave tool has been used to generate highly realistic ultrasound images, to study the sources of image degradation~\cite{pinton2011erratum,pinton2011effects} and to understand the principles behind new imaging methods, such as short lag spatial coherence imaging~\cite{dahl2012harmonic,pinton2014spatial}. It has also been used to simulate how elements that are blocked by ribs can degrade image quality~\cite{Jakovljevic2017blocked} and to describe multiple scattering due to microbubble contrast agents \cite{joshi2017iterative}. This simulation approach has been used to successfully model human transcranial focused ultrasound therapy of the brain. The direct propagation approach along with the three dimensional nonlinear propagation wave physics has been used with time-reversal acoustics to correct for individual skull morphology and to determine heat deposition  \cite{fink1992time,pernot2007vivo,pinton2011effects,soulioti2019super,pinton2012direct}.  {In the context of TBI, the Fullwave tool has been used to design high frame-rate imaging sequences to observe brain motion during traumatic impacts~\cite{Espindola2017}.}

 {\subsection{Brain motion during impact}}
Measuring brain motion during an injurious impact, which lasts only tens of milliseconds, has been a persistent challenge in biomechanics. High frame-rate ultrasound has the right combination of imaging frequency and soft-tissue contrast to sample tissue motion in the nonlinear elastic regime~\cite{catheline2003observation,espindola2017flashfocus}. High frame-rate imaging techniques in combination with adaptive correlation-based tracking algorithms~\cite{pinton2014adaptive} have been used to observe that, the shear waves generated by an impact to the brain can develop into destructive shear shock waves~\cite{Espindola2017}. 
At the shock front the acceleration is amplified by a factor of up to 8.5, i.e. a 35$g$ wave develops into a 300$g$ wave deep inside the brain, which may be the primary mechanism responsible for diffuse axonal injuries. This complex brain motion is governed by nonlinear viscoelastic wave physics. It has been shown that for a {linearly-polarized} plane wave excitation, where the particle motion is in the axis orthogonal to the plane of wave propagation, brain motion can be described analytically by a cubically nonlinear version of Burgers' equation~\cite{Espindola2017}. Although this motion can be quite complex, it has a clear odd harmonic signature in the frequency spectrum, i.e. only the third, fifth, seventh, etc. multiples of the fundamental frequency are generated by the nonlinear propagation. In other words, to capture the nonlinear brain motion, for a planar shear wave one must also capture the odd harmonic spectrum. 

This is a somewhat more challenging problem than what is required for shear wave elastography. In shear wave elastography, the Young's modulus is estimated from the shear wave speed \cite{sarvazyan1998shear} and only the bulk of the energy at the fundamental frequency must be measured to obtain this estimate. However, for a nonlinear shear wave, the energy at the higher harmonics must also be correctly estimated. This high spectral sensitivity requirement implies that the shape of the waveform must be correctly characterized. Furthermore, the signal at the harmonics is typically {40 dB} lower than the fundamental i.e. the components that must be detected to measure this waveform have a low amplitude compared to the bulk energy~\cite{Espindola2017}. 

 {\subsection{Incorporation of brain motion for FDTD simulations}}

The motion generated by a shear wave, e.g. in elastography, can be accurately modeled by finite difference simulations\cite{pinton2017continuous}. Displacements that were much smaller than the grid size (up to $\lambda/6000)$ have been previously modeled in Fullwave simulations, using an impedance flow method~\cite{pinton2017continuous}. This method consist {ed} of  generating scatterers in the acoustic simulation field, with each scatterer being composed of two spatial pixels. Then, motion  {could} be represented by redistributing the impedance between the two constituent pixels of each scatterer. In this manner, continuous values of subresolution displacements  {were} modeled using a coarse simulation grid size ($\lambda/15$). 

The two-pixel impedance flow method was applied to scatterer distributions in a uniform background medium. However, tissue has heterogeneous acoustical structures from embedded connective tissue, blood vessels, skin/fat layers and, in the case of brain, white matter, grey matter and cerebrospinal fluid.  {Here, we present a) a generalization of the impedance flow method that is applicable to any heterogeneous tissue maps with an arbitrary distribution of impedance, b) validation of the method by imposing known shear shock wave displacements in acoustical maps of the human brain.}

 {Unlike the smooth shear waves generated by a low amplitude acoustic radiation force, the shear shock waves observed during brain injury have sharp profiles that are more challenging to represent on finite grids. The shear shock wave displacements therefore represent a challenging scenario to evaluate the impedance flow method in terms of representing the shock front characteristics such as rapid displacements and the unique odd harmonic spectral content.}

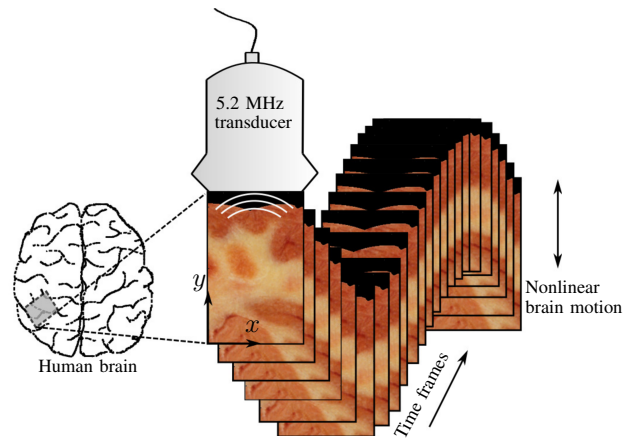
\begin{figure}[ht]
\centering
\setlength{\unitlength}{0.475\textwidth}
\begin{picture}(0.05,0.6)(0.03,0.3)

\put(-0.4,0.28){\includegraphics[height=0.31\textwidth,trim=120 0 0 0,clip]{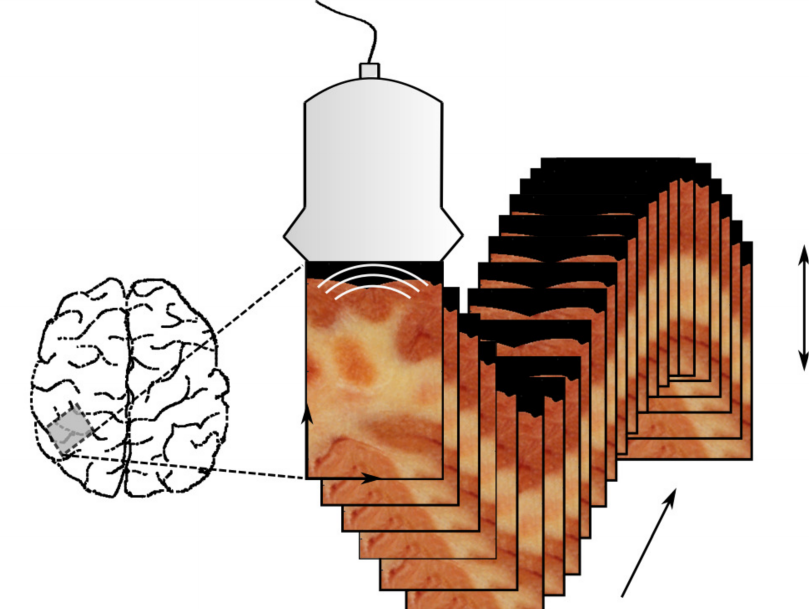}}

\put(-0.06,0.78){\scriptsize  5.2 MHz}
\put(-0.062,0.75){\scriptsize transducer}
\put(0.41,0.5){\scriptsize Nonlinear }
\put(0.41,0.47){\scriptsize brain motion}
\put(-0.33,0.38){\scriptsize Human brain}

\put(-0.03,0.44){ \color{black} \textbf{$x$}}
\put(-0.11,0.51){ \color{black} \textbf{$y$}}

\begin{turn}{62} 
\put(0.35,-0.07){\scriptsize Time frames}
\end{turn}

\end{picture}

\caption{Illustration of the ultrasound imaging plane and  direction of shear shock wave motion.  Shear wave displacements were imposed on the map of the brain along the $y$-axis. Ultrasound imaging simulations were performed in each of the displaced brain maps (time or frame axis). }

\label{fig:schematic}
\end{figure}
\begin{figure}[h]

  \begin{center}
\tikzstyle{block} = [draw, fill=white, rectangle, minimum height=3em, minimum width=6em]
\tikzstyle{sum} = [draw, fill=white, circle, node distance=1cm]
\tikzstyle{input} = [coordinate]
\tikzstyle{output} = [coordinate]
\tikzstyle{pinstyle} = [pin edge={to-,thin,black}]
\begin{tikzpicture}[auto, node distance=15mm,>=latex']
\footnotesize
    \node [block,text width=30mm](map){Generate acoustical map of brain from Visible Human};
    \node [block,text width=30mm, right of=map,node distance=40mm](scatterer){Generate subresolution scatterer map};
    \node [block,text width=30mm, below of=map,node distance=20mm](impflow){ {Use impedance flow method to calculate new impedance values in brain map to implement displacement }};
      \node [block,text width=30mm, below of=impflow, node distance=20mm](ppm){ {Use PPM solver to simulate the shear wave displacements to be imposed}};
          \node [block,text width=30mm, below of=scatterer,node distance=30mm](fullwave){ {Simulate ultrasound imaging sequence using Fullwave for each displaced brain map }};
          \draw[->](ppm)--(impflow);
             \draw[->](map)--(impflow);
             \draw[->](scatterer)--(impflow);
  
     \node [block,text width=30mm, below of=fullwave](beamform){Beamform simulated RF data};
    \node [block,text width=30mm, below of=beamform](track){Track motion with adaptive correlation-based algorithm};
          \draw[->](impflow)-|(fullwave);
          \draw[->](fullwave)--(beamform);
            \draw[->](beamform)--(track);
            
                \node [block,text width=30mm, below of=track](compare){ {Compare measured motion with known imposed brain motion}};
                            \draw[->](track)--(compare);

\end{tikzpicture}
  \end{center}
  \caption{Schematic of the acoustic map generation, ultrasound simulation, beamforming, and shear shock wave tracking in the brain.}
  \label{fig:appraoch_summary}
\end{figure}
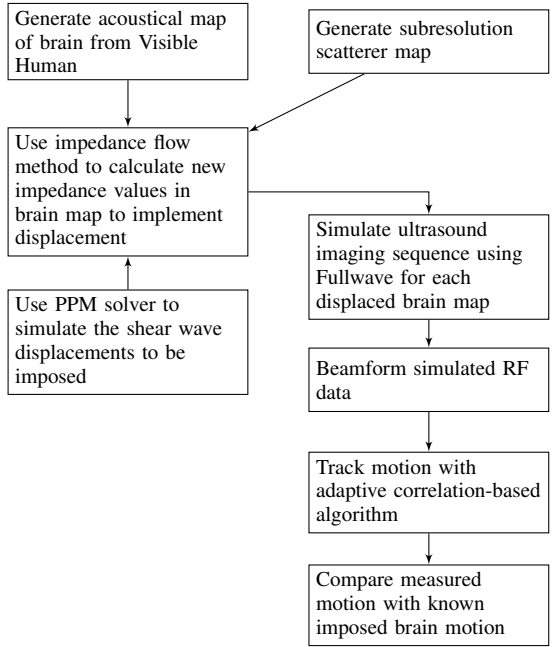

To simulate ultrasound propagation in the brain while it is moving due to the complex shear shock wave deformation from a traumatic impact (Fig. \ref{fig:schematic}), we propose here a generalized impedance flow method that describes continuous subresolution displacements in heterogeneous acoustical maps.
Optical maps of the human brain from the Visible Human project \cite{visiblehuman1998} are converted into acoustical maps that segment white matter, grey matter, cerebrospinal fluid and include subresolution scatterers. The proposed impedance flow method is validated using these tissue maps for the a full range of subresolution displacements, between 0 and 1 pixels. Then, shear shock wave displacements are calculated with a custom piecewise parabolic method (PPM) based solver that was previously calibrated to model nonlinear viscoelastic motion due to the shock wave propagation in brain-mimicking gelatin phantoms \cite{Tripathi2017}  {(discussed in III, E).} This motion is imposed on the acoustical maps  of brain tissue. The imposed displacement is validated by tracking beamformed RF data backscattered from the reference and displaced brain maps. This process is summarized schematically in Fig.~\ref{fig:appraoch_summary}. It is shown that the proposed method accurately captures the nonlinear motion of the brain in terms of the shock profile, harmonic spectrum, and acceleration, including acceleration at the shock front, which are parameters that are relevant to traumatic brain injury.

\section{Methods}

\subsection{Acoustical maps of the human brain for Fullwave simulations}



\begin{figure*}[h]
\centering
\setlength{\unitlength}{\textwidth}
\begin{picture}(1,0.25)(0,-0.0)

\put(0.1,-0.3){\includegraphics[height=12cm,trim=5 75 80 120,clip]{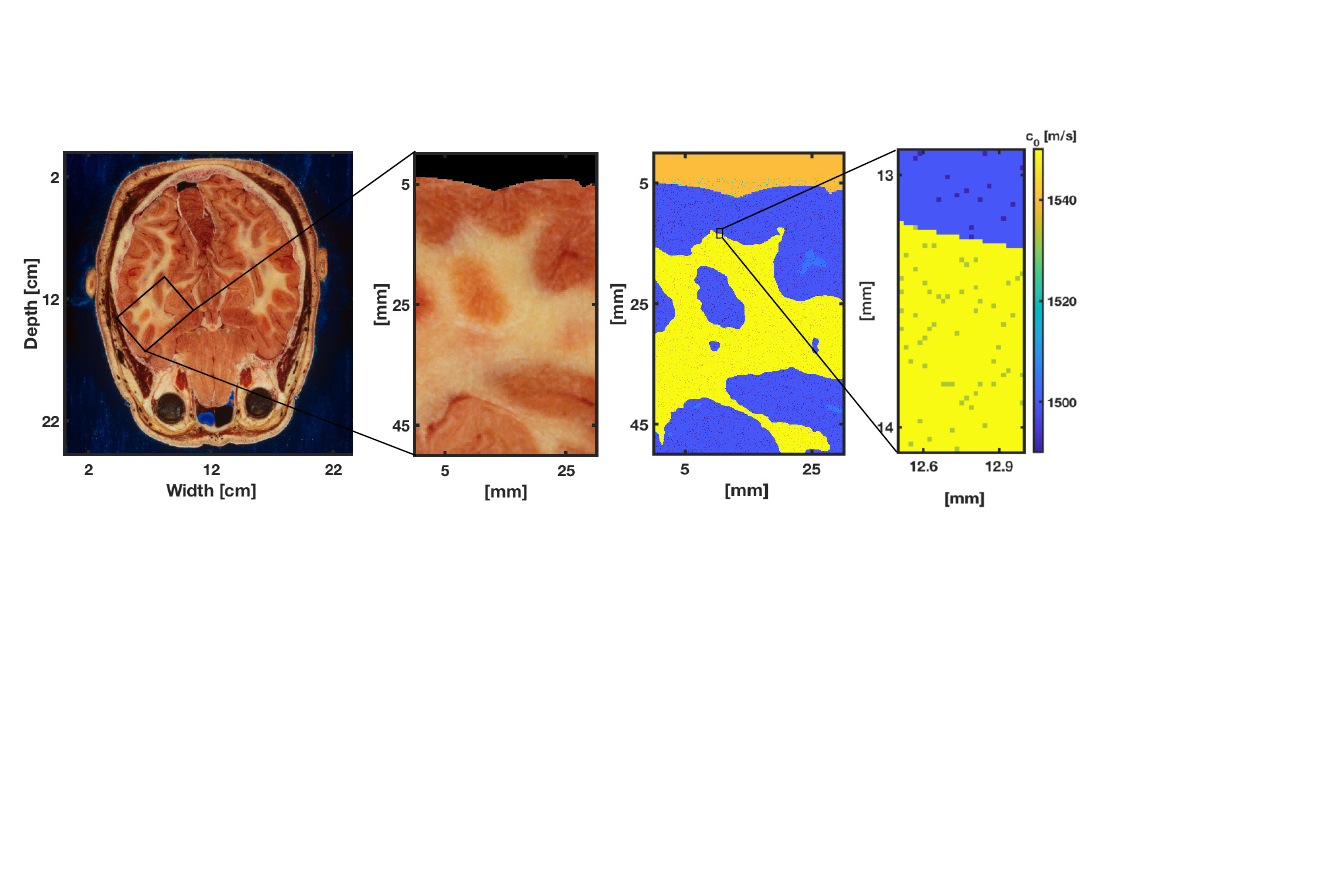}}


\put(0.33,0.225){\scriptsize \color{white}\textbf{(a)}}
\put(0.51,0.225){\scriptsize \color{white} \textbf{(b)}}
\put(0.69,0.225){\scriptsize \textbf{(c)}}
\put(0.82,0.225){\scriptsize \color{white} \textbf{(d)}}

\end{picture}

\caption{a) Image of a human brain slice in RGB format obtained from the Visible Human project dataset \cite{visiblehuman1998}. b) Fullwave simulation domain interpolated to grid size of 0.018 mm (${\lambda/16}$). c) Sound speed map created using RGB thresholding and median filtering. d) Pixel-sized scatterers imposed and assigned sound speed values 2.5\% lower than the local tissue sound speed. }

\label{fig:brain}
\end{figure*}
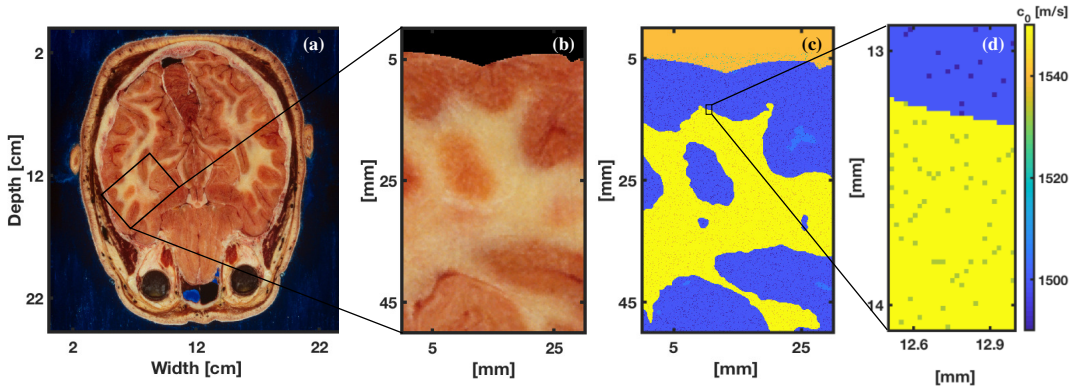

Photographic cryosections from the Visible Human data set were used to construct anatomically accurate acoustical maps of the human brain.  {An arbitrary 3$\times$5 cm region of the brain was chosen to include multiple tissue interfaces, from the transverse plane at the level of the eye (Fig.~\ref{fig:brain}(a,b)). The size of the imaging region was chosen based on an active aperture size of 3~cm.} This region was linearly interpolated from an optical resolution of 330~${\rm \mu m}$ to 18~${\rm \mu m}$ in order to obtain a grid with 16 points per wavelength, relative to the center frequency of the ultrasound imaging pulse (5.2~MHz).  Different tissue types namely, gray matter, white matter and cerebrospinal fluid were detected using RGB threshold filters. First, the green channel of the optical image was extracted and the white matter regions were filtered out by setting a lower bound  threshold of 135 on the image intensity scale. Similarly, the cerebrospinal fluid regions were separated from the gray matter regions using a filter with an upper bound intensity threshold of 65. Subsequently, a 12$\rm \times$12 pixel spatial median filter was applied to the thresholded image until continuous tissue boundaries were detected. Post the separation of tissue regions, discrete sound speed values of 1552.5 m/s, 1500 m/s and 1504.5 m/s were respectively assigned to white matter, gray matter and cerebrospinal fluid regions (Fig.~\ref{fig:brain}(c)). Finally, the region in the imaging field outside of the brain boundary was assigned a water path between the transducer and the brain surface, with a sound speed of 1540 m/s. 
The resulting acoustical map was heterogeneous only in terms of sound speed. A constant density of  1000~${\rm kg/m^3}$ was assigned to the entire field. 
Considering a constant density, impedance values in the map were modified during displacement imposition, by only changing the sound speed values, based on the relationship ${\rm z=\rho c}$. 

 {To generate ultrasound images of the brain map with fully developed speckle, pixel-sized scatterers were distributed throughout the field using a uniform probability density function, to generate an average density of $>$12 scatterers per resolution area (Fig.~\ref{fig:brain}(c)). The scatterers (\num{0.21e6} in total) constituted 4.8\% of the imaging field which consisted of \num{4.4e6} spatial cells. The average density of scatterers per resolution area was uniform however, their spatial distribution was random. The scatterers were assigned sound speed values that were 2.5\%  lower than the local tissue map sound speed.} This is similar to previous simulations of ultrasound imaging using the Fullwave simulation tool~\cite{pinton2009heterogeneous}. These scatterers are too small to see on the full map, but they are visible in a zoomed-in portion (Fig.~\ref{fig:brain}(d)).

\subsection{Ultrasound imaging simulations with Fullwave}
 {Fullwave simulations were performed in 2D in the lateral plane. Note that the imaging region was restricted to within the brain without including the skull, to only incorporate the nonlinear displacement model of the brain as simulated from the PPM solver (described in II,D). The plane of imaging is chosen to be orthogonal to the direction of propagation of the shear wave, in order to impose a uniform displacement throughout the field.} 

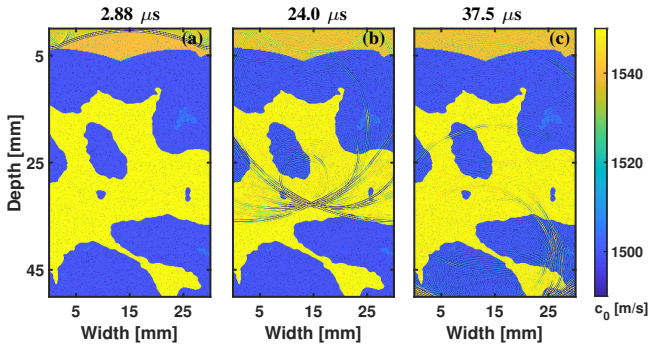
\begin{figure}[h]
\centering
\setlength{\unitlength}{0.45\textwidth}
\begin{picture}(1,0.5)(0,-0.0)
\put(-0.02,-0.04){\includegraphics[height=4.2cm,trim=0 0 80 0,clip]{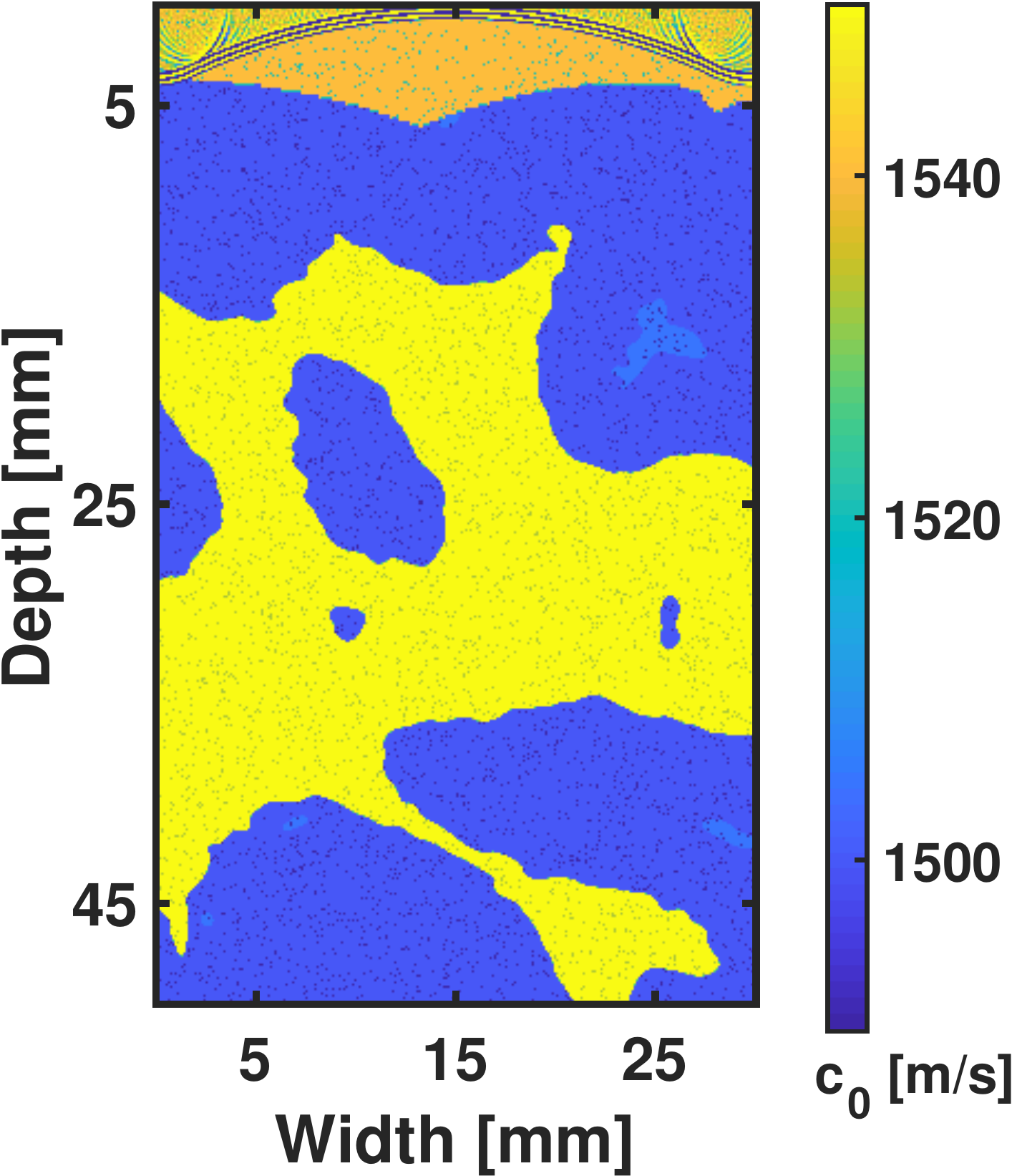}}

\put(0.34,-0.04){\includegraphics[height=4.2cm,trim=60 0 80 0,clip]{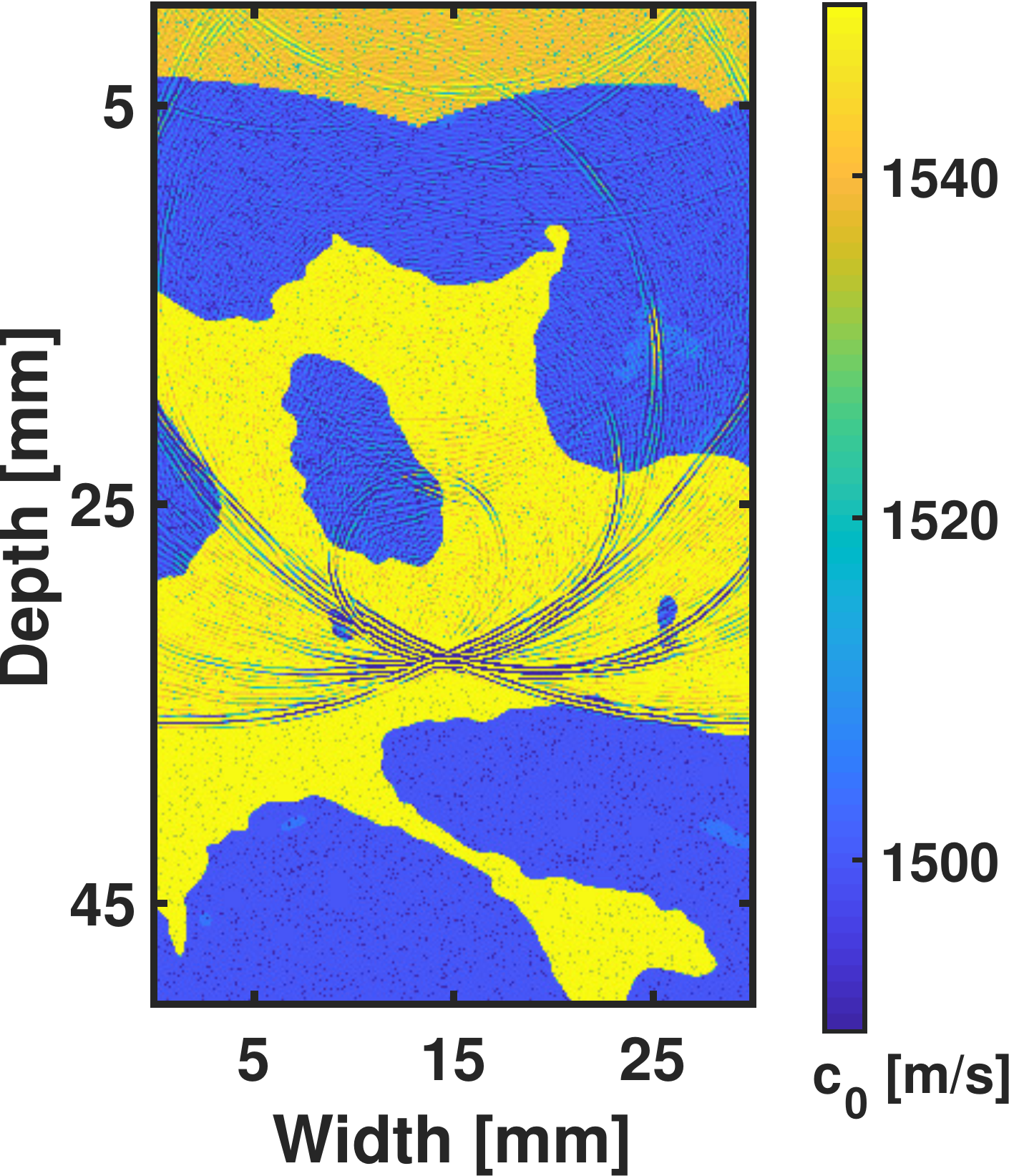}}

\put(0.63,-0.04){\includegraphics[height=4.2cm,trim=60 0 0 0,clip]{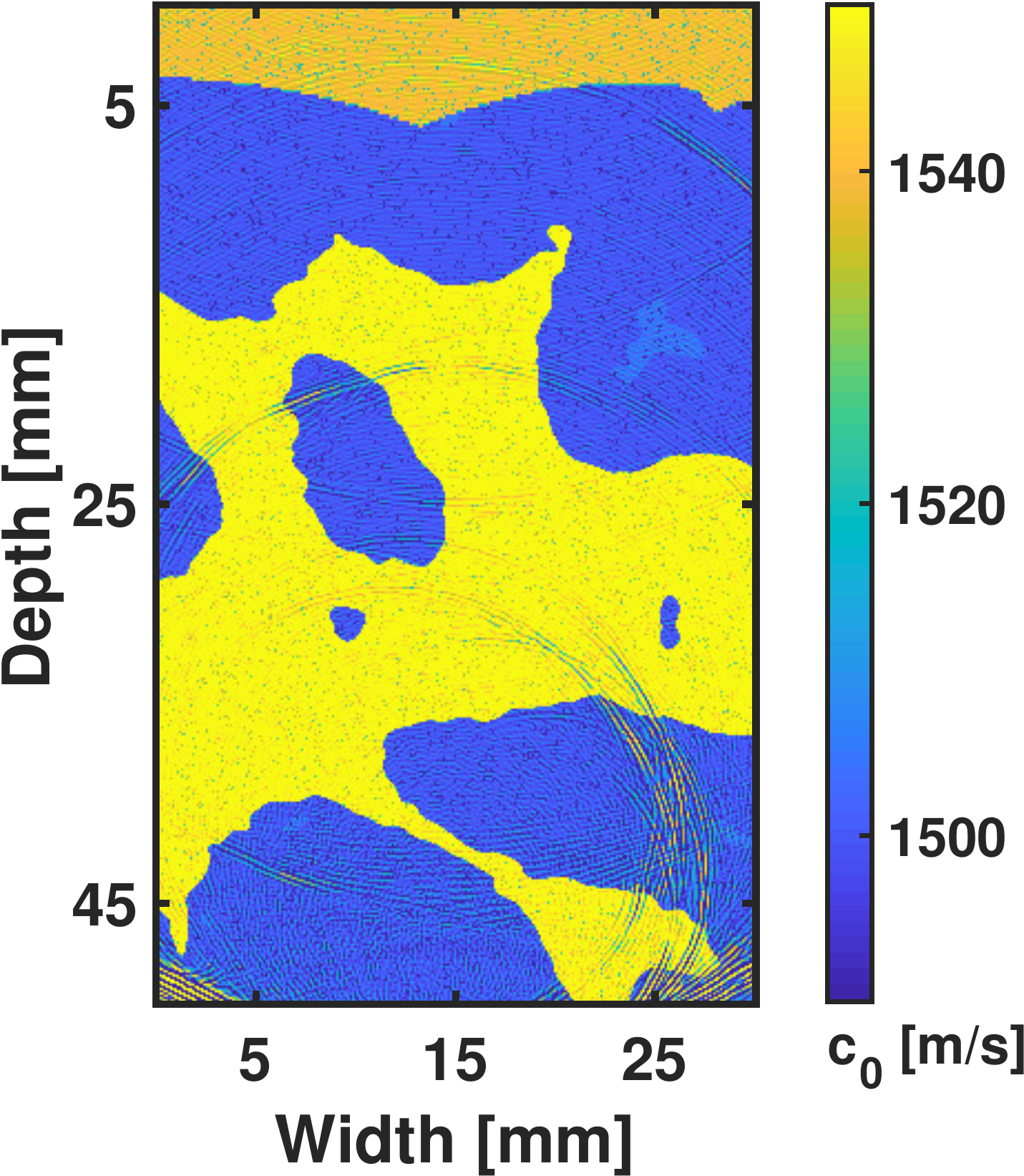}}

\put(0.13,0.479){\scriptsize \textbf{2.88 $\mu$s}}
\put(0.43,0.479){\scriptsize \textbf{24.0 $\mu$s}}
\put(0.71,0.479){\scriptsize  \textbf{37.5 $\mu$s}}

\put(0.26,0.44){\scriptsize \color{black}{\textbf{(a)}}}
\put(0.55,0.44){\scriptsize \color{black}{\textbf{(b)}}}
\put(0.85,0.44){\scriptsize  \color{black}{\textbf{(c)}}}

\end{picture}
\caption{Three snapshots in time of the Fullwave simulation modeling the L7-4 clinical array with a 5.2 MHz 2-cycle imaging pulse emitted from a 3 cm aperture. The pressure is overlaid on the speed of sound map of the brain map. a) The initial pulse illustrating the focal profile b) The pulse at the focal position. at a depth of 3.12 cm. c) The backscattered wave.}

\label{fig:propagation}
\end{figure}

 {The simulated imaging sequences modeled the ATL Philips L7-4 transducer with an active aperture of 3 cm and a 5.2~MHz center frequency. The transducer emitted a 2 cycle pulse (Fig.~\ref{fig:propagation}(a)) focusing along the mid-line axis of the transducer at a depth of 3.12 cm (Fig.~\ref{fig:propagation}(b)).}  {The elevation focus was not accounted for in the 2D imaging field.} The Courant-Friedrichs-Lewy (CFL) condition was set to 0.4, which at 16 points per wavelength is equivalent to a the time discretization of 4.8~ns.  {The numerically simulated backscattered synthetic RF data (hereon referred to as ``RF data")} (Fig.~\ref{fig:propagation}(c)) was recorded at 208~MHz over a duration of 82~${\rm \mu s}$ and subsequently beamformed using a conventional delay-and-sum beamforming algorithm. The acoustic propagation in the medium was assumed to be linear and non-attenuating. 
By measuring the arrival time of the reflection from a scatterer placed at the focus, the average sound speed of the brain field was calculated to be 1537 m/s. An imaging simulation was performed for each realization or frame of the acoustical maps, to generate one ultrasound image, per brain position (Fig.~\ref{fig:schematic}). 

\begin{subsection}{Subresolution displacement in the heterogeneous tissue map using a generalized impedance flow method}

The continuous subresolution shear motion of the brain was imposed along the depth direction, or $y$-axis in Fig.~\ref{fig:schematic}. For displacements that were an integral multiple of the grid size, the brain map was shifted by a discrete number of depth pixels on the existing grid. This is the trivial displacement scenario.  However, this approach cannot be applied to  displacements that are less than a pixel, without reducing the grid size to  {be less than or equal to} the displacement increment. Therefore, sub-pixel brain displacements were implemented using a generalized form of the impedance flow method. By the proposed method, subresolution displacements were generated in the acoustical map shown in Fig.~\ref{fig:brain}(c) as illustrated in  Fig.~\ref{fig:scat}(a,b), which shows a zoomed-in section of the reference and displaced brain sound speed maps at a gray-white matter tissue interface with two subresolution scatterers. 

 {Using the proposed impedance flow method, the reference and displaced maps shown in Fig.~\ref{fig:scat}(a) and (b) are obtained by imposing a displacement of 0.5 pixel and 0.7 pixel respectively (along the direction indicated using arrows in Fig.~\ref{fig:scat}), to the brain map shown in Fig.~\ref{fig:brain}(c-d) consisting of pixel-sized scatterers.} To impose a fractional displacement $\gamma$ by the proposed method, the impedance value in each spatial pixel is modified according to:


\vspace{-0.25cm}
\begin{eqnarray}
z_i'=(1-\gamma)~z_i+\gamma~z_{i-1} 
\label{proposed}
\end{eqnarray}
where, $z_i$ and $z_{i}'$ are respectively, the impedance values at the $i$th pixel location, prior to and post the displacement implementation. The variable $\gamma$  has values between $0$ and $1$, which correspond to no displacement and a one pixel displacement, respectively. 


 {For example, according to this formulation, for a 0.5 pixel displacement, the impedance at an interface of media 1 and 2 is modified as $(z_{1}+z_{2})/2$ thereby generating a one pixel thick boundary between the two media with the average value of impedance (indicated as `interface' in Fig.~\ref{fig:scat}(a)). Using the same formulation the  scattering elements in the field consist of two pixels with equal values of impedance (the sound speed values are indicated next to the scatterer pixels in Fig.~\ref{fig:scat} in m/s). Likewise, for a 0.7 pixel displacement, the impedance at the boundary is modified as $(0.3~z_{1}+0.7~z_{2})/2$ (indicated as `interface' in Fig.~\ref{fig:scat}(b)). Here, each scattering element consists of two pixels with different impedance values. The analytical interframe displacement between the maps in Fig.~\ref{fig:scat}(a) and (b) is 0.2 pixel. Therefore the expected time shift between their corresponding acoustical backscatter signals is 1 time sample, for the chosen CFL value of 0.4. The imposed displacement is verified as discussed in III A. }

 {For comparison, the reference and displaced maps shown in Fig.~\ref{fig:scat}(c) and (d) were obtained by displacing the brain map shown in Fig.~\ref{fig:brain}(c-d) by 0.5 pixel and 0.7 pixel respectively, by the scatterer-based impedance flow method as implemented in homogenous maps in \cite{pinton2017continuous}. By assuming each scatterer to be composed of two pixels, impedance is transferred between the constituent pixels to describe motion, while keeping the total scatterer impedance as conserved. The scatterer-based impedance flow approach renders identical impedance values as the generalization presented in Eqn.~\ref{proposed}, only for the case where tissue interfaces are absent. Note that the impedance values at the scatterers in Fig.~\ref{fig:scat}(c) and (d) are identical to those in Fig.~\ref{fig:scat}(a) and (b), respectively.  
However, the background maps are treated as stationary and thereby impedance values at tissue interfaces are unchanged (Fig.~\ref{fig:scat}(c) and (d)). 
This method is resolution limited given that the tissue map motion is rounded down to zero.}

Displacement imposed by the proposed generalized method (Fig.~\ref{fig:scat}(a-b)) is estimated and compared to that imposed by the scatterer-based impedance flow method (Fig.~\ref{fig:scat}(c-d)) in III A.



\begin{figure}[h]
\centering
\setlength{\unitlength}{0.475\textwidth}
\begin{picture}(1,0.48)(0.02,0.4)

\put(0.0,0.41){\includegraphics[height=0.26\textwidth,trim=70 180 240 60,clip]{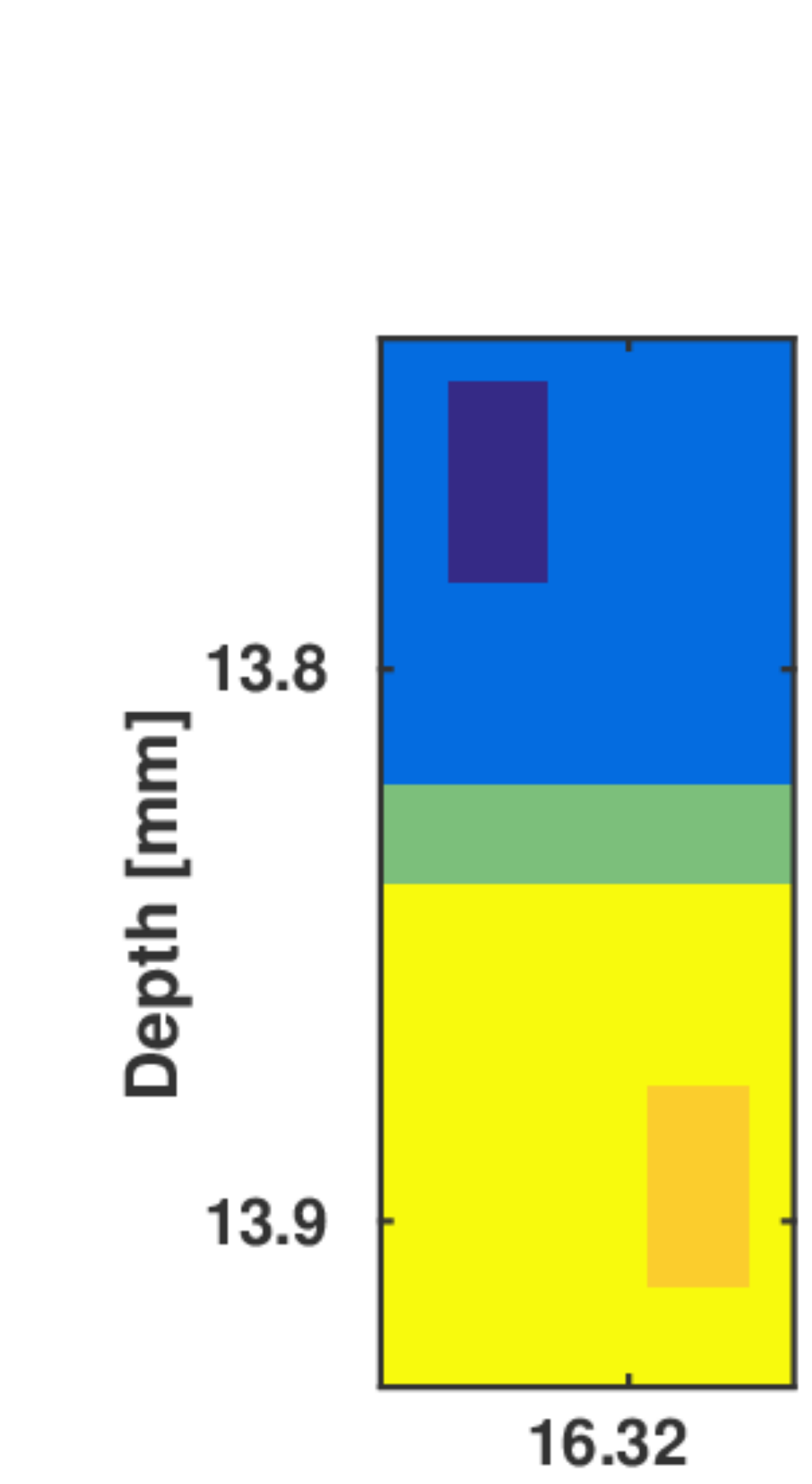}}
\put(0.325,0.41){\includegraphics[height=0.26\textwidth,trim=210 180 240 60,clip]{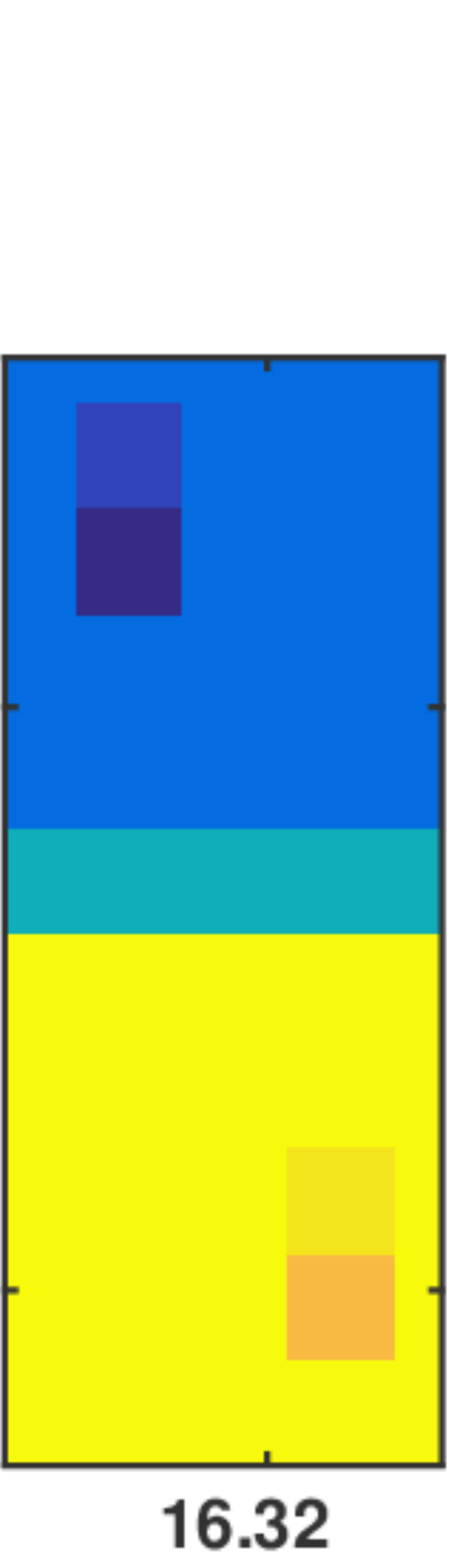}}

\put(0.275,0.7){\vector(0,-1){0.1}}
\put(0.46,0.7){\vector(0,-1){0.1}}

\put(0.52,0.41){\includegraphics[height=0.26\textwidth,trim=210 180 240 60,clip]{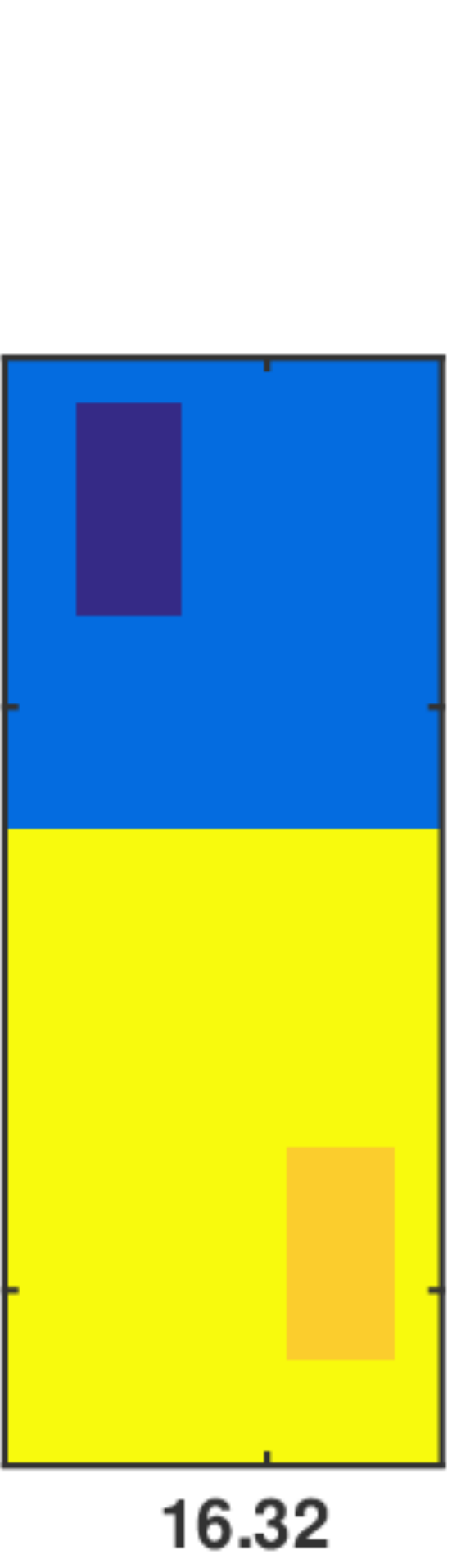}}
\put(0.7,0.41){\includegraphics[height=0.251\textwidth,trim=200 170 0 60,clip]{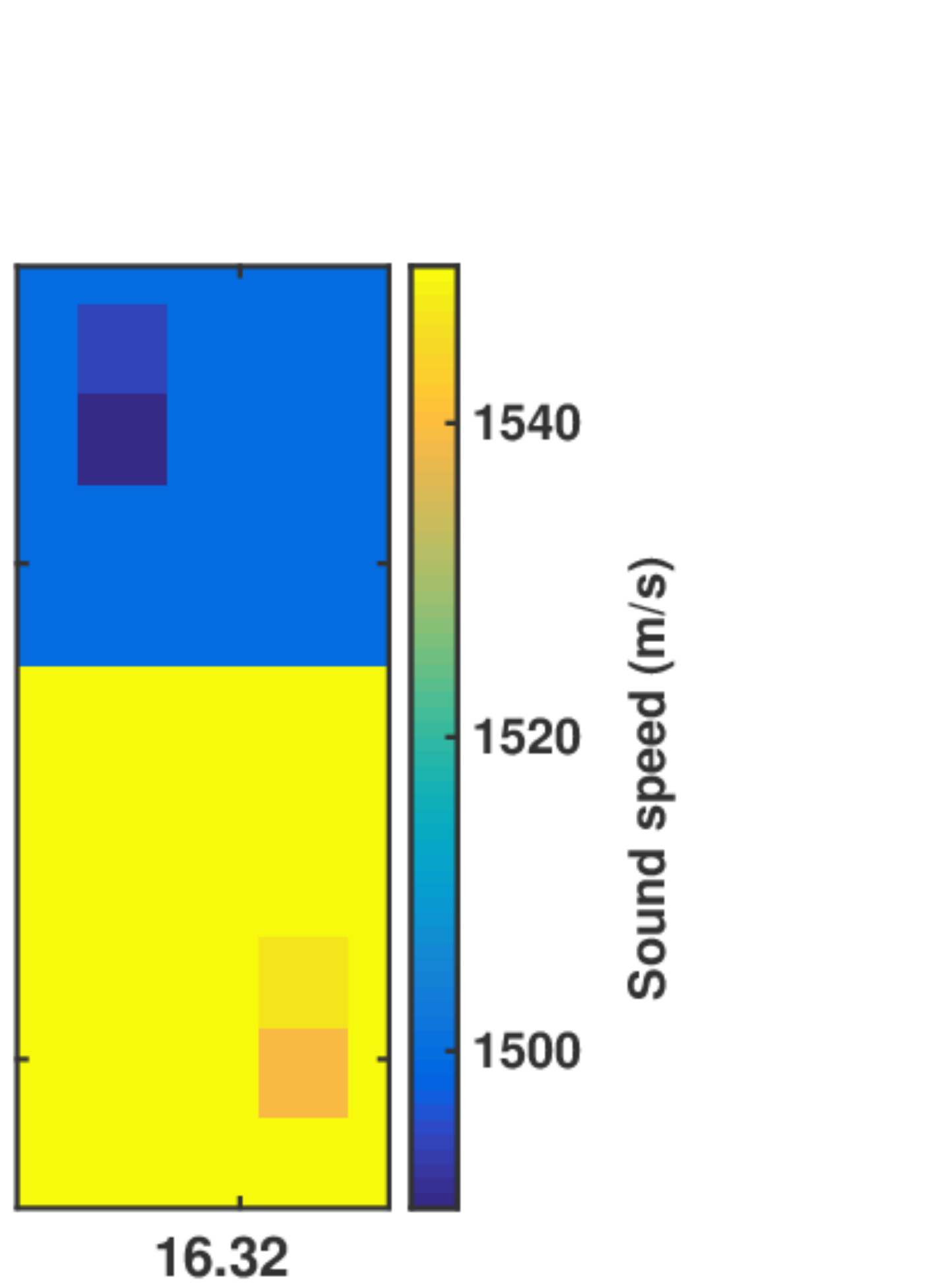}}

\put(0.64,0.7){\vector(0,-1){0.1}}
\put(0.83,0.7){\vector(0,-1){0.1}}

\put(0.08,0.22){\includegraphics[height=0.085\textwidth,trim=50 0 100 610,clip]{fig5a.pdf}}
\put(0.47,0.22){\includegraphics[height=0.085\textwidth,trim=50 0 100 610,clip]{fig5b.pdf}}

\put(0.205,0.75){\scriptsize \color{black} 1491}
\put(0.205,0.785){\scriptsize \color{black} 1491}
\put(0.18,0.53){\scriptsize \color{black} 1543}
\put(0.18,0.49){\scriptsize \color{black} 1543}

\put(0.39,0.75){\scriptsize \color{black} 1494}
\put(0.39,0.785){\scriptsize \color{black} 1487}
\put(0.36,0.53){\scriptsize \color{black} 1547}
\put(0.36,0.49){\scriptsize \color{black} 1539}

\put(0.585,0.75){\scriptsize \color{black} 1491}
\put(0.585,0.785){\scriptsize \color{black} 1491}
\put(0.56,0.53){\scriptsize \color{black} 1543}
\put(0.56,0.49){\scriptsize \color{black} 1543}

\put(0.775,0.75){\scriptsize \color{black} 1494}
\put(0.775,0.785){\scriptsize \color{black} 1487}
\put(0.74,0.53){\scriptsize \color{black} 1547}
\put(0.74,0.49){\scriptsize \color{black} 1539}

\put(0.15,0.46){\scriptsize \color{black} (a)}
\put(0.34,0.46){\scriptsize \color{black} (b)}
\put(0.53,0.46){\scriptsize \color{black} (c)}
\put(0.72,0.46){\scriptsize \color{black} (d)}

\put(0.16,.84){\scriptsize Reference}
\put(0.34,.84){\scriptsize Displaced}
\put(0.54,.84){\scriptsize Reference}
\put(0.73,.84){\scriptsize Displaced}
\put(0.15,.64){\scriptsize Interface}
\put(0.33,.64){\scriptsize Interface}
\end{picture}

\caption{ {Using the proposed generalized form of the impedance flow method, an interframe displacement of 0.2 pixels was imposed along the direction indicated using arrows, between the a) reference and the b) displaced maps. Impedance values are modified at the subresolution scatterers as well as at tissue interfaces (sound speed values shown at scatterer locations). (c-d) In contrast, the tissue maps were displaced in a resolution-limited manner, by treating the background tissue as stationary. Note that the scatterers are the same as in (a) and (b). }}

\label{fig:scat}
\end{figure}

\end{subsection}

\begin{subsection}{Shear shock displacements constructed by  Piecewise Parabolic Method (PPM) simulation}

{ {The nonlinear shear wave displacements that were imposed in the brain map (Fig.~\ref{fig:brain}(c)) were calculated using a PPM simulation tool,}  {previously validated using shear shock wave experiments~\cite{Tripathi2017,Espindola2018measurement}} This simulation tool numerically solves a Burgers-like equation  {in 1D}, with a cubic nonlinear term instead of a quadratic term\cite{zabolotskaya2004modeling}, and is written as:
\begin{equation}
    \frac{\partial v}{\partial z} = \frac{\beta}{c_t^3}v^2\frac{\partial v}{\partial \tau} +  {c\left(\mathcal{F}(v)\right) + \alpha\left(\mathcal{F}(v)\right).}    \label{shearshock}
\end{equation}
where $v$ is the particle velocity orthogonal to the direction of propagation $z$, and $\tau=t-z/c_t$ is the retarded-time (where the time-axis $t$ moves with the wave at the linear shear speed $c_t$). The nonlinear parameter $\beta=\frac{3\gamma}{2\mu}$, where $\mu$ and $\gamma$ are the linear and nonlinear elastic constants, respectively.  { The second and third terms respectively denote  dispersion and attenuation  which are imposed using an operator splitting approach using the Fourier spectrum of $v$, as denoted by $\mathcal{F}(v)$.}  

 {Equation \eqref{shearshock} is numerically resolved using PPM, a fourth order finite volume method\cite{Colella1984}.} In finite volume methods, the domain is discretized into cells/volumes and a piecewise-polynomial, in our case quadratic, is used to approximate the state-variable in each cell. 
They are designed to conserve the in- and out- flux of state-variable like mass, momentum etc.  inside each cell/volume \cite{Leveque2002-a}. This ensures that the entropy of the system is constant and is verified using the Rankine-Hugoniot jump conditions \cite{Smoller2012}. This is an important test for numerical solvers with ability to simulate shock waves, otherwise it could predict a shock with a wrong speed. 


 {Attenuation and dispersion are incorporated in the frequency domain. At each spatial point, the Fourier transform of the numerically solved velocity signal is multiplied by frequency-dependent attenuation and dispersion law. Then the modified Fourier spectrum is converted to time domain to obtain the velocity signal at the subsequent spatial point.}

\begin{figure}[H]
\setlength{\unitlength}{0.475\textwidth}
\begin{picture}(1,0.78)(0,0)
\centering
\put(0.02,0.42){\includegraphics[width=0.225\textwidth,trim=0 0 0 0,clip]{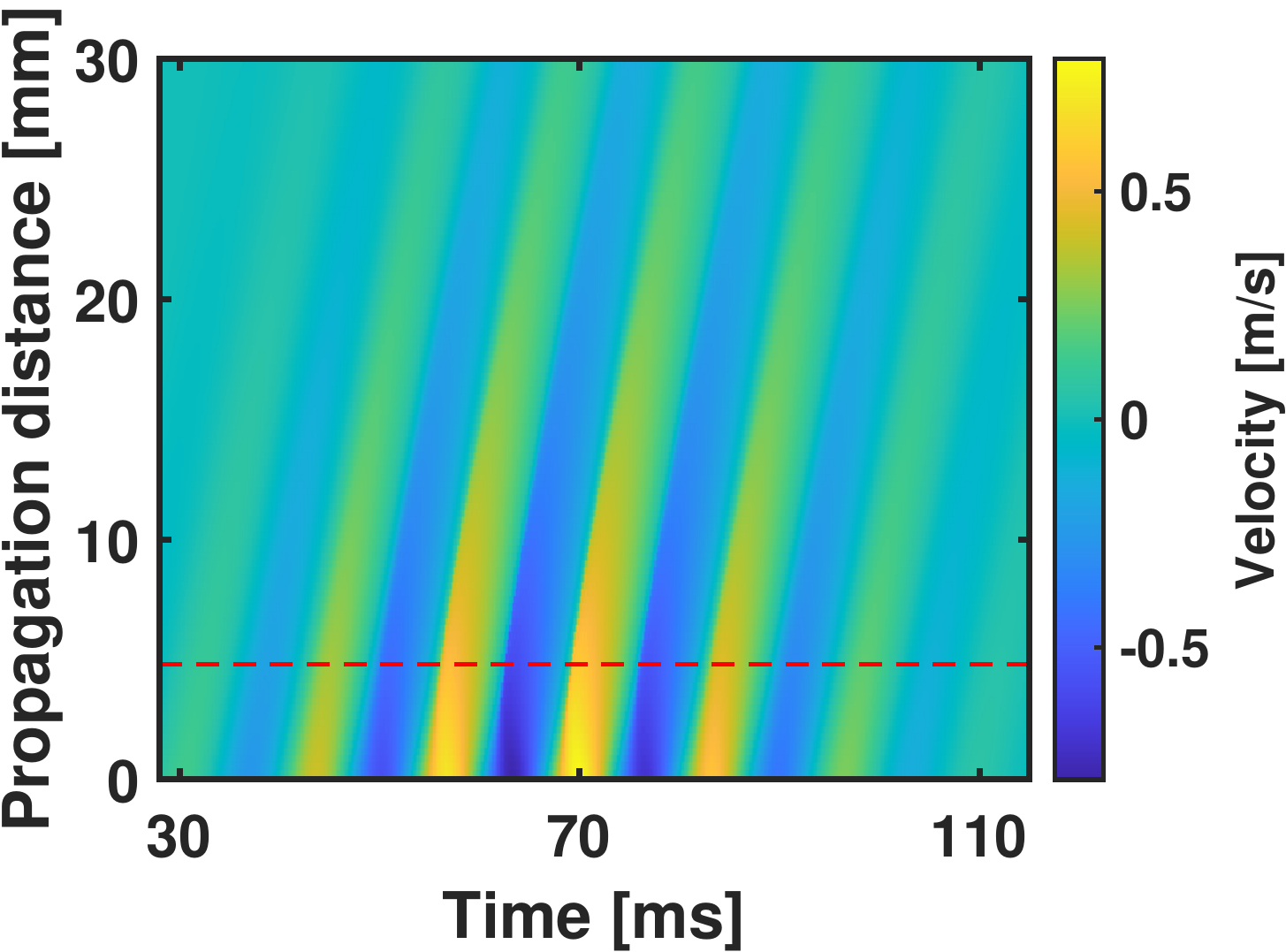}}
\put(0.52,0.42){\includegraphics[width=0.23\textwidth,trim=0 0 0 0,clip]{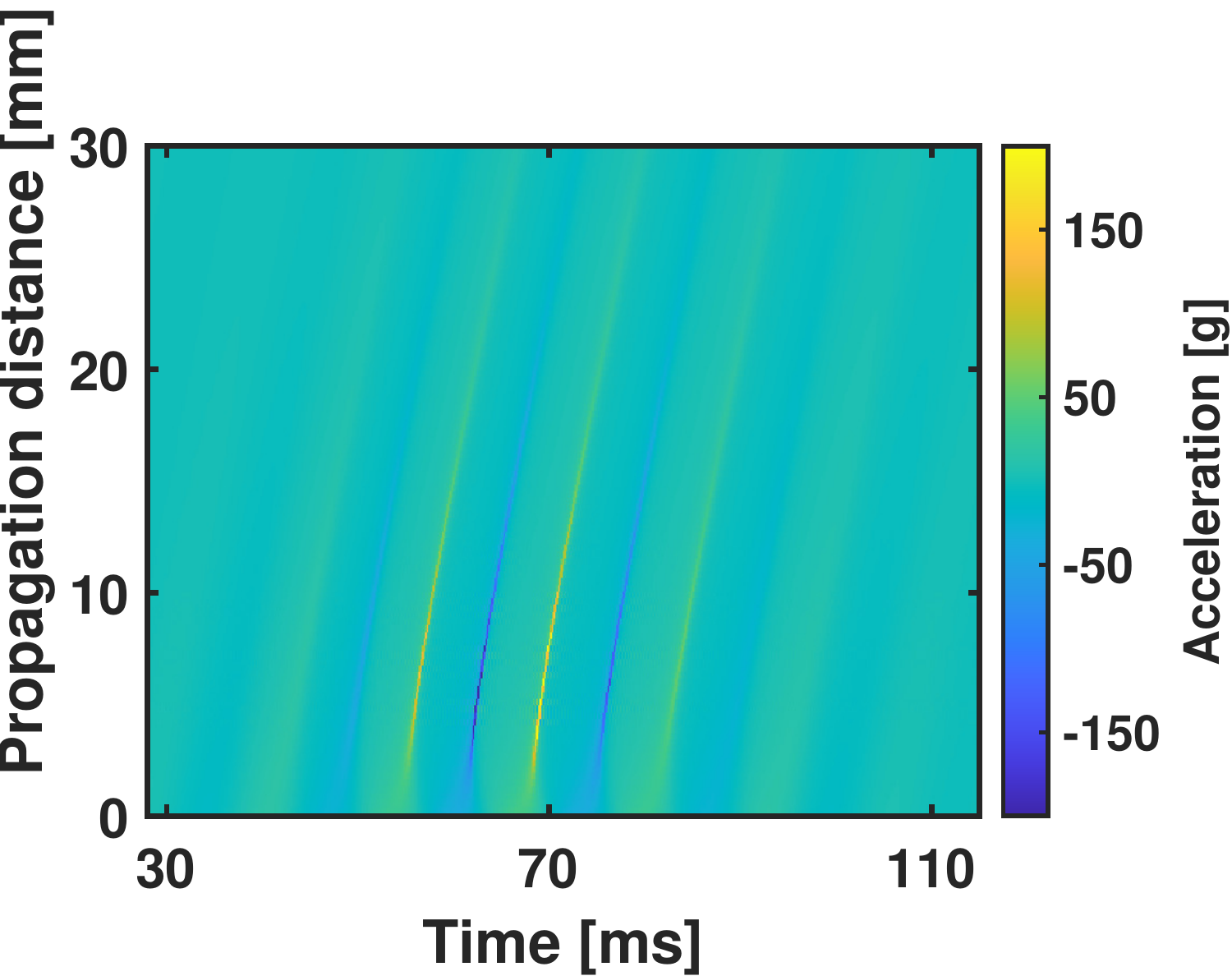}}

\put(0.0,0.022){\includegraphics[width=0.22\textwidth,trim=0 0 0 0,clip]{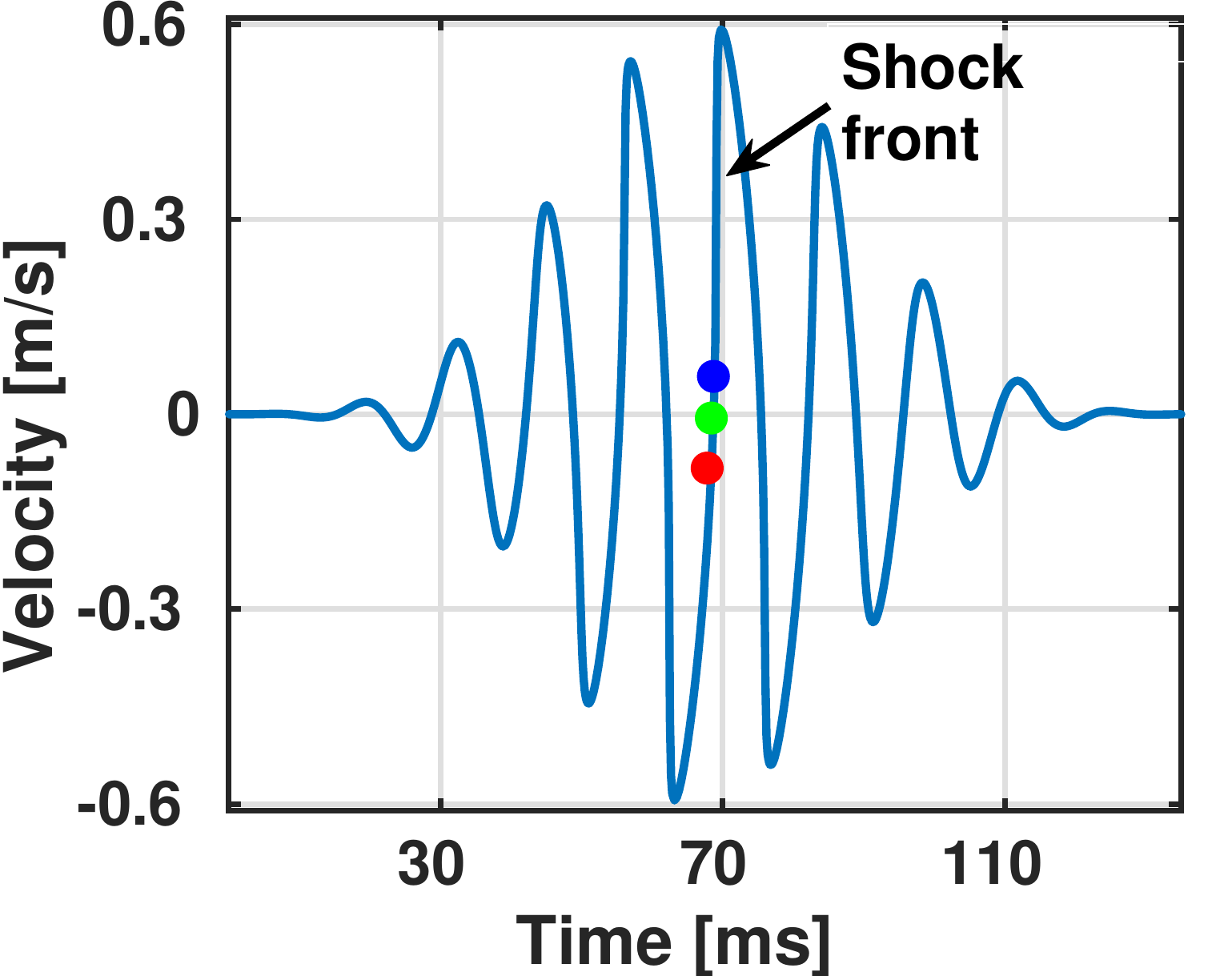}}
\put(0.5,-0.0){\includegraphics[width=0.14\textwidth,trim=0 0 115 0,clip]{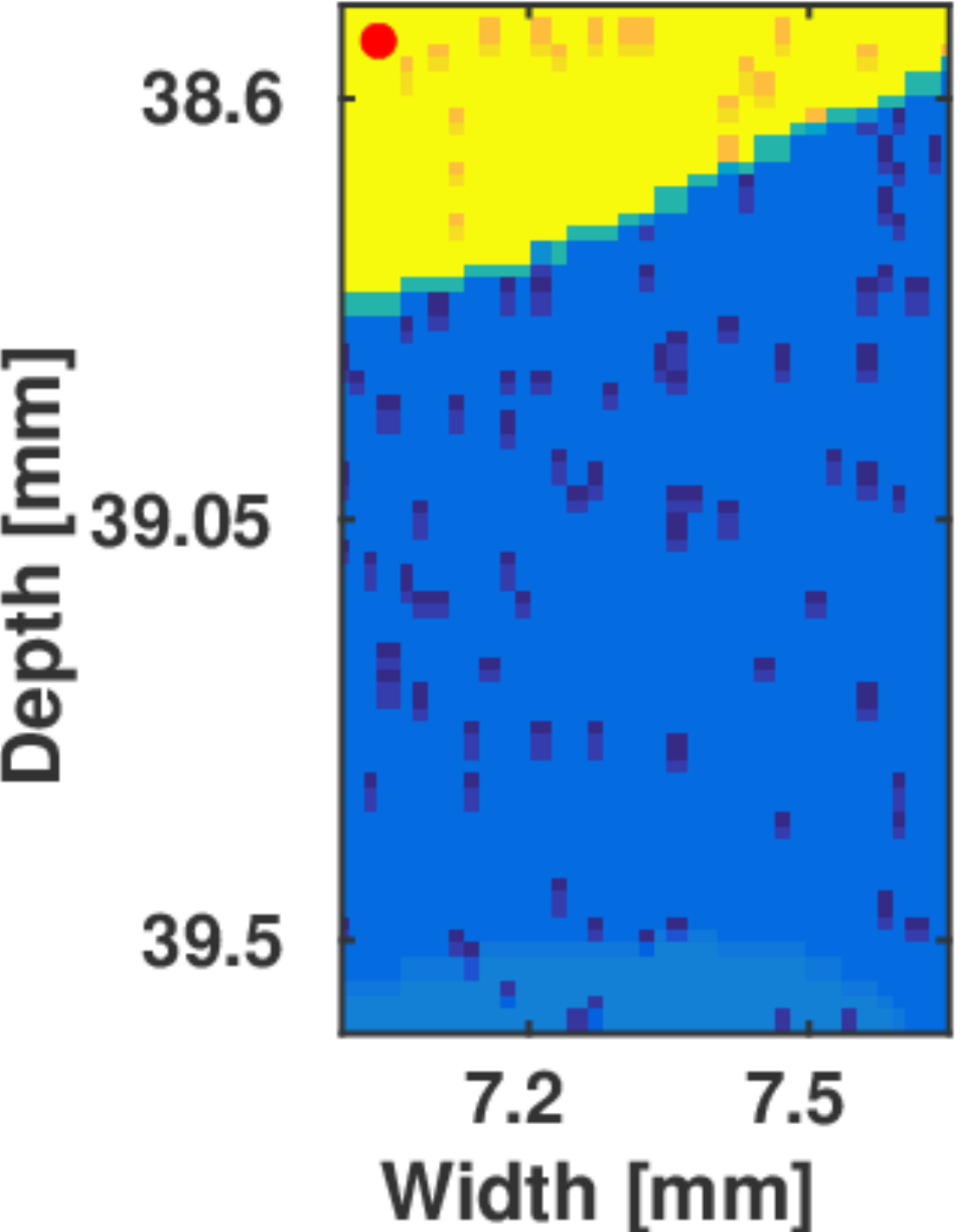}}
\put(0.65,0.06){\includegraphics[width=0.095\textwidth,trim=113 64 115 0,clip]{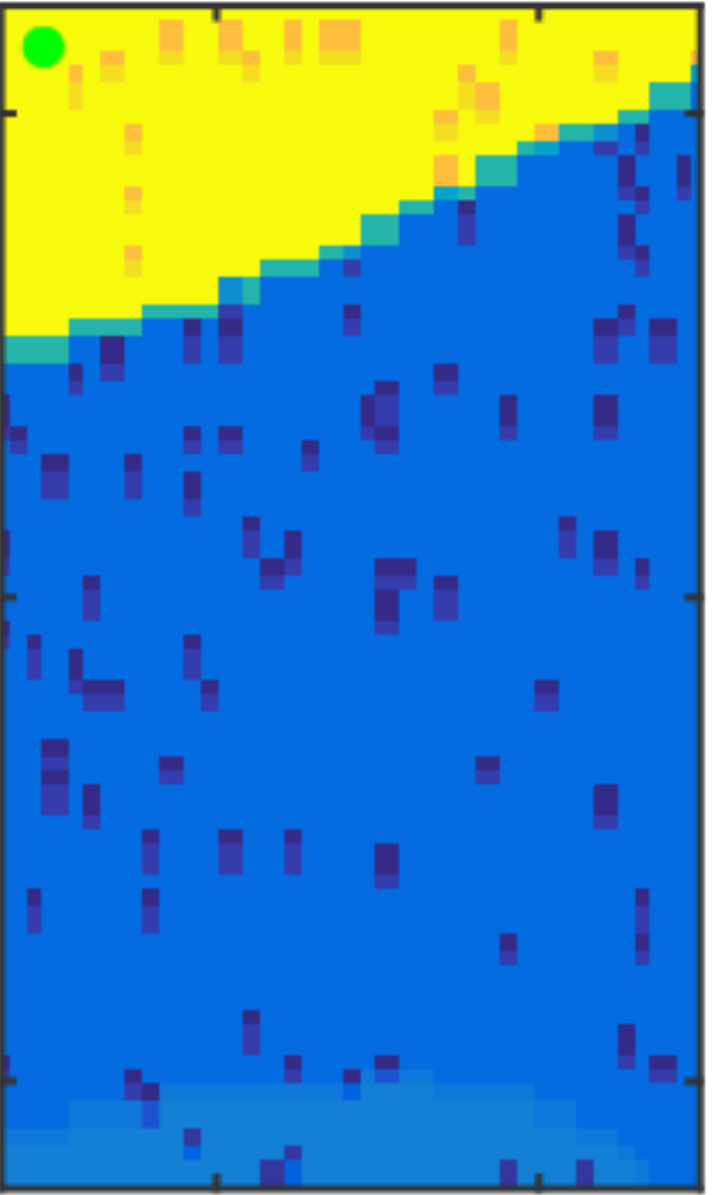}}
\put(0.7,0.09){\includegraphics[width=0.14\textwidth,trim=113 64 0 0,clip]{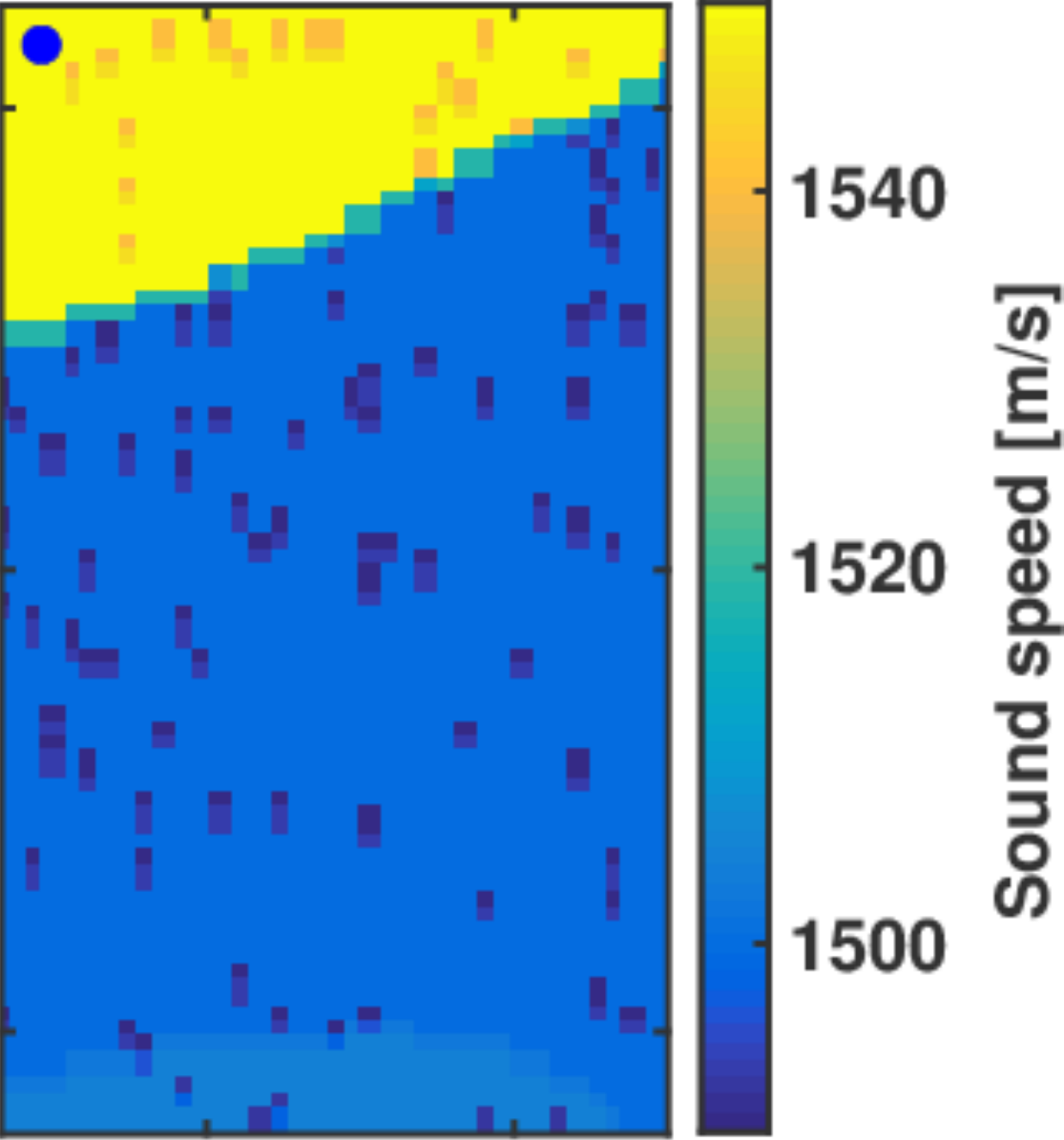}}

\put(0.7,0.09){\vector(1,0){0.07}}
\put(0.7,0.09){\vector(0,1){0.07}}
\put(0.78,0.09){\textbf{x}}
\put(0.69,0.176){\textbf{y}}

\put(0.36,0.71){\scriptsize (a)}
\put(0.85,0.71){\scriptsize (b)}
\put(0.405,0.35){\scriptsize (c)}
\put(0.845,0.375){\scriptsize (d)}

\end{picture}
\caption{ {Shear wave simulation was performed in 1D using the PPM solver in a homogeneous medium with mechanical properties of brain as measured from ex vivo experiments \cite{Espindola2017}. Spatio-temporal maps of the numerically solved shear wave velocity and  acceleration are shown in a) and b), respectively. c) The shear shock velocity waveform at 5.1~mm from the source was chosen as the input for imposing shear displacements on the brain maps. d) For each time point shown in c), the corresponding displacement was imposed on the acoustical brain map along \textit{y} (also indicated in Fig.~\ref{fig:schematic}) using the proposed impedance flow method.}}



\label{fig:PPM}
\end{figure}
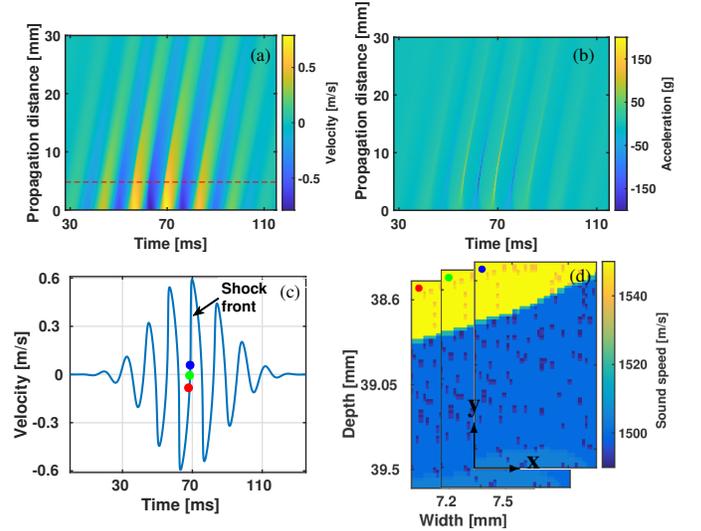
 {Although the simulated shear wave displacements were imposed in a heterogenous acoustical brain map (Fig.~\ref{fig:brain}(c)), the PPM shear wave simulations were performed for a homogeneous medium with the same average physical parameters as brain because the shear wavelength ($\sim$2 cm) is much larger than the acoustical wavelength. In other words, the small spatial scale of the acoustical variations were assumed to be irrelevant for shear wave propagation. The simulation medium was set to have a shear wave speed of $c_t=2.14 ~{\rm m/s}$ at 75 Hz and  nonlinear parameter $\beta=13$ with attenuation $\alpha(\omega)=0.21\omega ^{0.89}$ Np/m and constant dispersion $c(\omega) = 2.14$  m/s. }   {The physical parameters of brain are based on  measurements performed in {\it ex vivo} shear wave experiments as reported in \cite{Espindola2017}. } 

 {The numerical solution of the PPM simulation is shown in terms of velocity and acceleration in Fig.~\ref{fig:PPM}(a) and (b), respectively. In Fig.~\ref{fig:PPM}(a) and (b),  the shear source is located at a propagation distance of 0~mm. The fully developed shear shock waveform at a propagation distance of 5.1mm (Fig.~\ref{fig:PPM}(c), corresponding to dashed red line in Fig.~\ref{fig:PPM}(a)) was chosen for displacement implementation. The chosen velocity waveform consists of 1160 samples, each of which was converted to displacements (in depth pixels) with respect to the first frame. Each of these displacements were imposed on the acoustical brain map shown in Fig.~\ref{fig:brain}(c). This steps generates 1160 brain maps, displaced according to the instantaneous shear wave velocity imposed at each time point (illustrated using the red, green and blue markers on the waveform in Fig.~\ref{fig:PPM}(c) and on the maps shown in Fig.~\ref{fig:PPM}(d)). }}

\end{subsection}

\begin{subsection}{Adaptive tracking}

 {The tracking algorithm was applied to one beamformed RF line located at the midline axis of the transducer. Each of the displaced brain maps generated an RF line. This is equivalent to an M-mode acquisition. Tracking could have been performed on full B-mode images but it would have increased the bookkeeping effort without changing the outcome of subsequent analysis that does not depend on lateral position in this case.}
A quality-weighted correlation-based tracking algorithm designed for shear shock wave propagation~\cite{pinton2014adaptive} was used to measure the displacements from the beamformed simulated RF data.  {  Although the magnitude of the  RF from the brain maps may vary weakly with displacement, depending on the impedance variation at tissue boundaries, only the phase shift is estimated here in order to measure displacement.} 

The tracking algorithm relies on a quality-weighted median filter that replaces low correlation  {displacement estimates} (below 0.75) with surrounding high quality estimates. The output of the median filter is then used as the initial guess for the subsequent displacement estimate iteration, which is performed over a smaller kernel and a smaller or the same search window. The known displacements imposed by the shear shock wave motion (Fig.~\ref{fig:PPM}(a)) were measured for each of the 1160 displaced fields (Fig.~\ref{fig:PPM}(b)). The beamformed RF data was interpolated by a factor of 3 along the depth dimension prior to the displacement tracking. The iterative displacement tracking was performed for kernel lengths varying from $10\lambda$ to $4\lambda$ and with a search window of 18 pixels in the first iteration and 2 pixels subsequently. The weighted median filter was sized 6 pixels in the depth dimension and 2 pixels in the time frame dimension.



\end{subsection}

\section{Results}

\subsection{Validation of the proposed impedance flow method}

\begin{figure}[h]
\setlength{\unitlength}{0.475\textwidth}
\begin{picture}(1,1.17)(0,0.15)
\centering

\put(0.04,1.1){\includegraphics[width=0.455\textwidth,trim=135 350 130 350,clip]{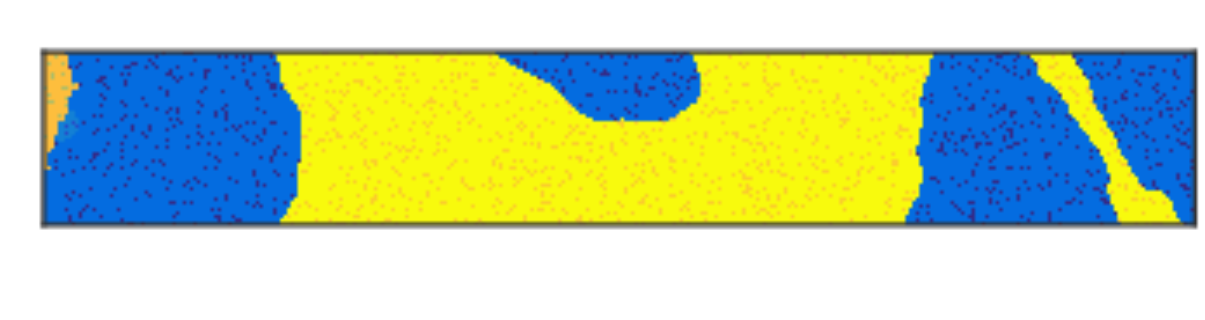}}

\put(-0.055,0.7){\includegraphics[width=0.54\textwidth,trim=30 0 30 15,clip]{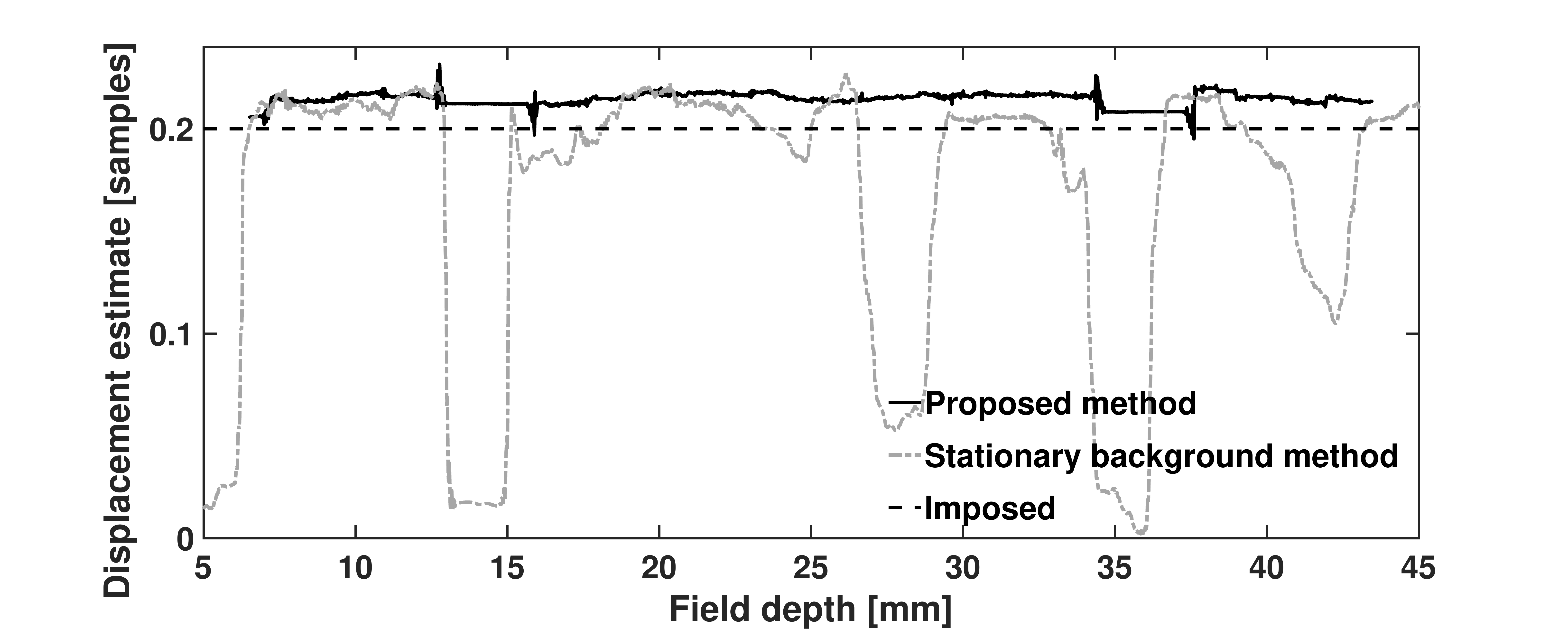}}
\put(-0.05,0.13){\includegraphics[width=0.535\textwidth,trim=30 0 30 20,clip]{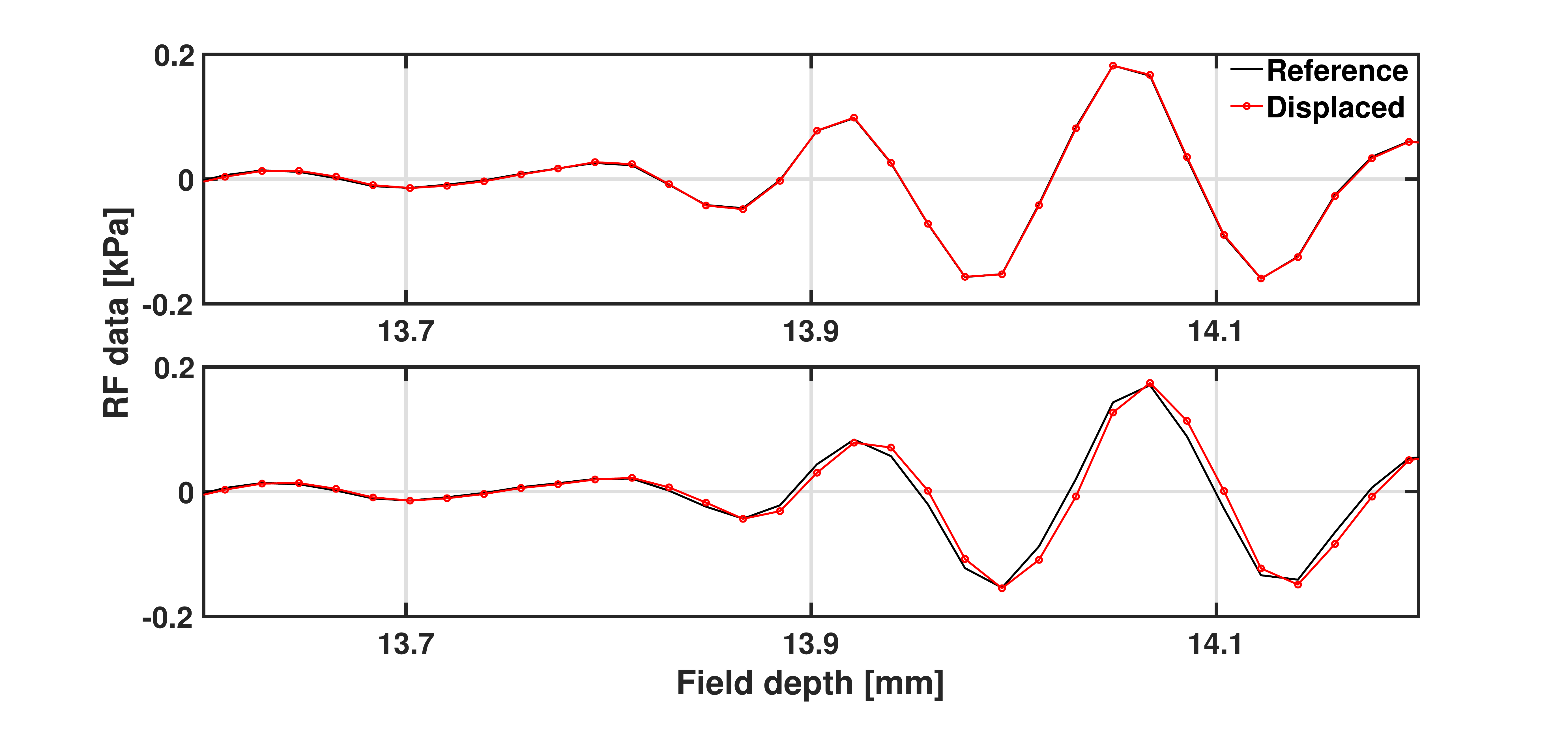}}
    
\put(0.1,1.23){\scriptsize \color{white}I-1}
\put(0.29,1.22){\scriptsize I-2}
\put(0.52,1.22){\scriptsize I-3}
\put(0.73,1.22){\scriptsize I-4}
\put(0.86,1.22){\scriptsize \color{white}I-5}
\put(0.95,1.22){\scriptsize \color{white}I-6}

\put(0.95,1.11){\scriptsize \textbf{\color{black}(a)}}

\put(0.94,0.47){\scriptsize \textbf{ {(b)}}}
\put(0.94,0.23){\scriptsize \textbf{ {(c)}}}

\end{picture}

\caption{a)  {Numerical} RF from Fullwave simulations were tracked for an imposed 0.2 pixel interframe displacement (dashed black line), by the proposed generalized impedance flow method (solid black line) and by the resolution-limited method (dashed gray line). The inset shows the field of view where the tracking is performed. The field of view is rotated by 90 degrees to register to the estimates shown in (a). The tissue interfaces in the field are indicated as I1 to I6. b) RF data zoomed-in at the depth location of a gray-white matter interface from maps displaced by the proposed method. c) RF data from maps displaced by the resolution-limited method.}

\label{fig:depth}
\end{figure}
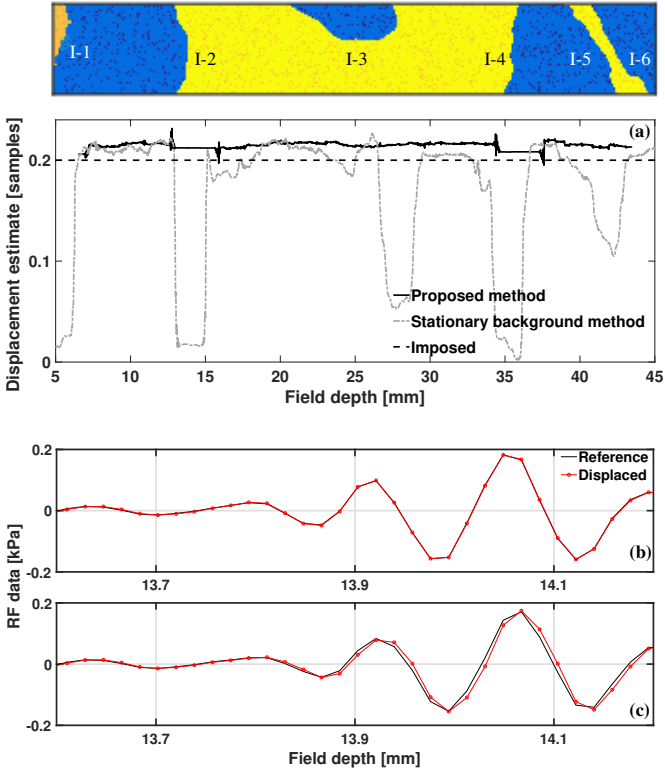

The beamformed RF from the Fullwave simulations of the L7-4 imaging transducer were compared for the cases where the scatterers and the tissue boundaries were displaced by 0.2 pixel using the proposed generalized impedance flow method and by the resolution-limited method. These two cases correspond to the tissue maps shown in Fig.~\ref{fig:scat}(a,b) and Fig.~\ref{fig:scat}(c,d), respectively. The RF data in Fig.~\ref{fig:depth}(b) and (c), is zoomed in at the depth location of the interface shown in Fig.~\ref{fig:scat}(a,b) and (c,d), respectively. There is a shift between the RF from the reference and displaced maps in the proposed case. However, there is no shift in the RF from the resolution-limited maps. This shows that the strong specular reflection from the tissue interface dominates the displacement estimate. Furthermore, this analysis shows that the approach of moving the scatterers while treating the background tissue map as stationary is insufficient to represent motion.

This analysis was then performed on the wider field of view (Fig.~\ref{fig:brain}(c)). The inset in Fig.~\ref{fig:depth}(a) shows the region where the tracking is occurring and it is rotated by 90 degrees so that it can be registered to the accompanying displacement estimates. These displacements were estimated along the mid-line axis of the transducer by using RF data that was beamformed using a sound speed value of 1537 m/s. Subsequently the RF data was interpolated by a factor of 3 and tracked using normalized cross correlation with a kernel length of $10\lambda$ (3 mm) and search window of 5 pixels to capture a range of 0.6 pixel displacements, post interpolation.

For the stationary background map (Fig.~\ref{fig:depth}, dashed gray line), the displacement estimates at the tissue interfaces (indicated as I1-I6 in Fig.~\ref{fig:depth}(a)), which should be 0.2 pixel, are much lower. The size of the region affected by this error is consistent with the kernel size. Furthermore, this result is consistent with the interpretation the specular reflection from the stationary tissue map dominates the displacement estimate. This is particularly evident at interfaces I-2 and I-4 where the interface orientation is orthogonal to the direction of wave propagation and is thus efficient at sending sound directly back to the transducer. 

In contrast, displacement estimates from the proposed impedance flow method  (Fig.~\ref{fig:depth}, solid black line) is able to accurately represent the displacement at the tissue boundaries in addition to the homogeneous scatterer regions. The error in the displacement estimate was averaged within a $10\lambda$ (3 mm) window at the focal depth (3.12 cm). It was calculated as 0.0143 spatial pixel, equivalent to $\lambda/1118$.  {The error is quantified in terms of both spatial pixels as well as the acoustical wavelength $\lambda$, given that the grid size ($\lambda/16$) is calibrated with respect to $\lambda$.}


\subsection{Bias and jitter}

The proposed generalized impedance flow method was then characterized in terms of bias and jitter. Bias was calculated as the average difference between the estimated and the analytical displacement from 50 different speckle realizations. The bias error provides information regarding the DC component. Jitter was calculated as the standard deviation of the error between the estimated and the analytical displacement. The analysis was performed for a range of displacements varying from -0.5 to +0.5 pixel. Positive displacements were imposed away from the transducer.  Each displacement value was characterized for 50 different realizations of the random scatterer distribution while retaining the same background brain tissue map, to obtain speckle-independent estimates. The focused RF data in each realization was beamformed using the same sound speed value of 1537 m/s and interpolated by a factor of 3. In each displaced field, the interframe displacement was estimated with respect to the reference field with no displacement. Estimates were obtained with three different kernel lengths namely $4\lambda$ (1.2 mm), $10\lambda$ (3 mm) and $50\lambda$ (15 mm), using a search window of 5 pixels. The error was calcuted by averaging the estimates within a $6\lambda$ window at the focal depth.

\begin{figure}[ht]
\centering
\setlength{\unitlength}{0.475\textwidth}
\begin{picture}(1,0.45)(0,0.02)

\put(-0.01,0.){\includegraphics[height=0.205\textwidth,trim=0 0 0 0,clip]{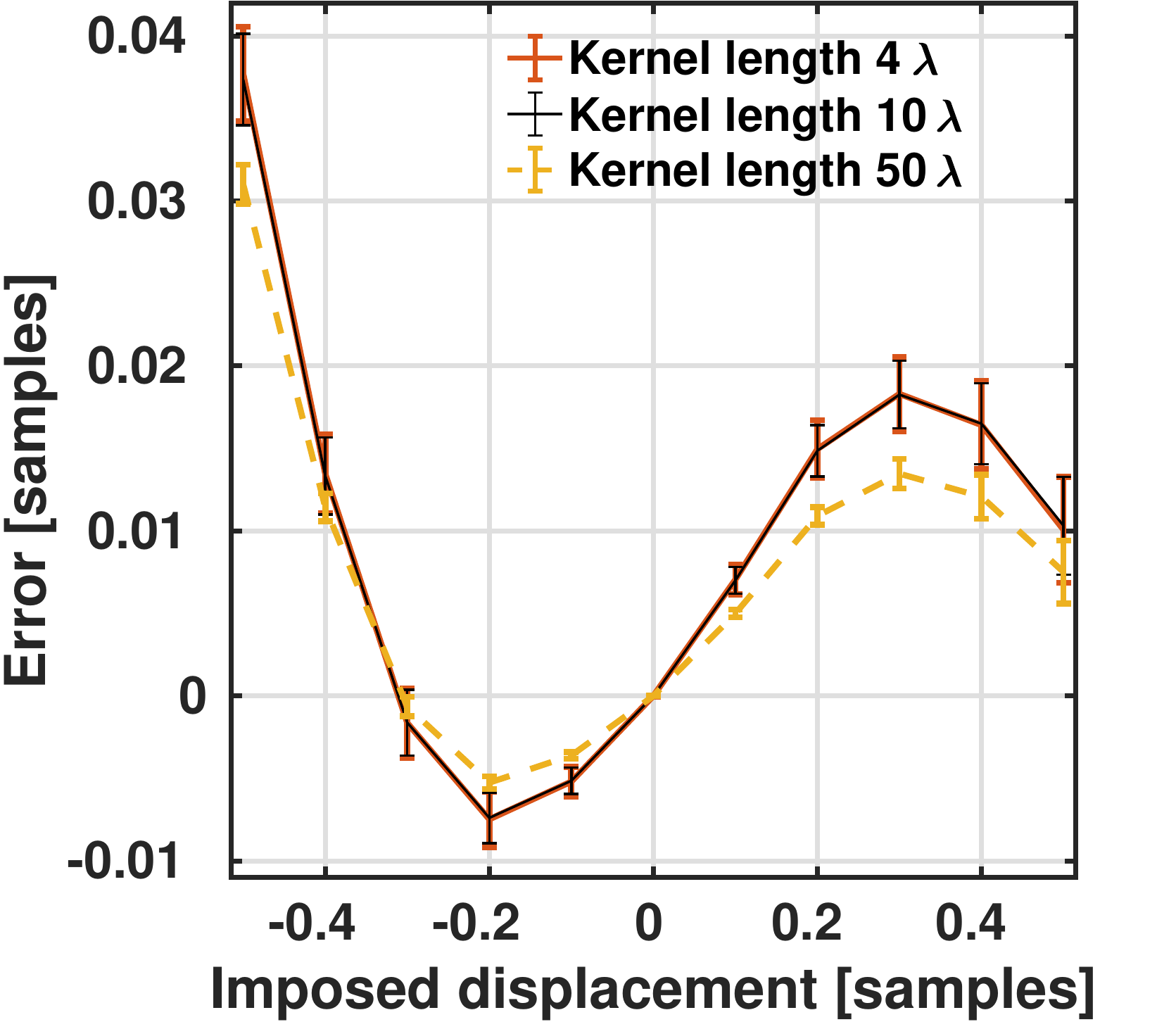}}
\put(0.512,0.){\includegraphics[height=0.205
\textwidth,trim=0 0 0 0,clip]{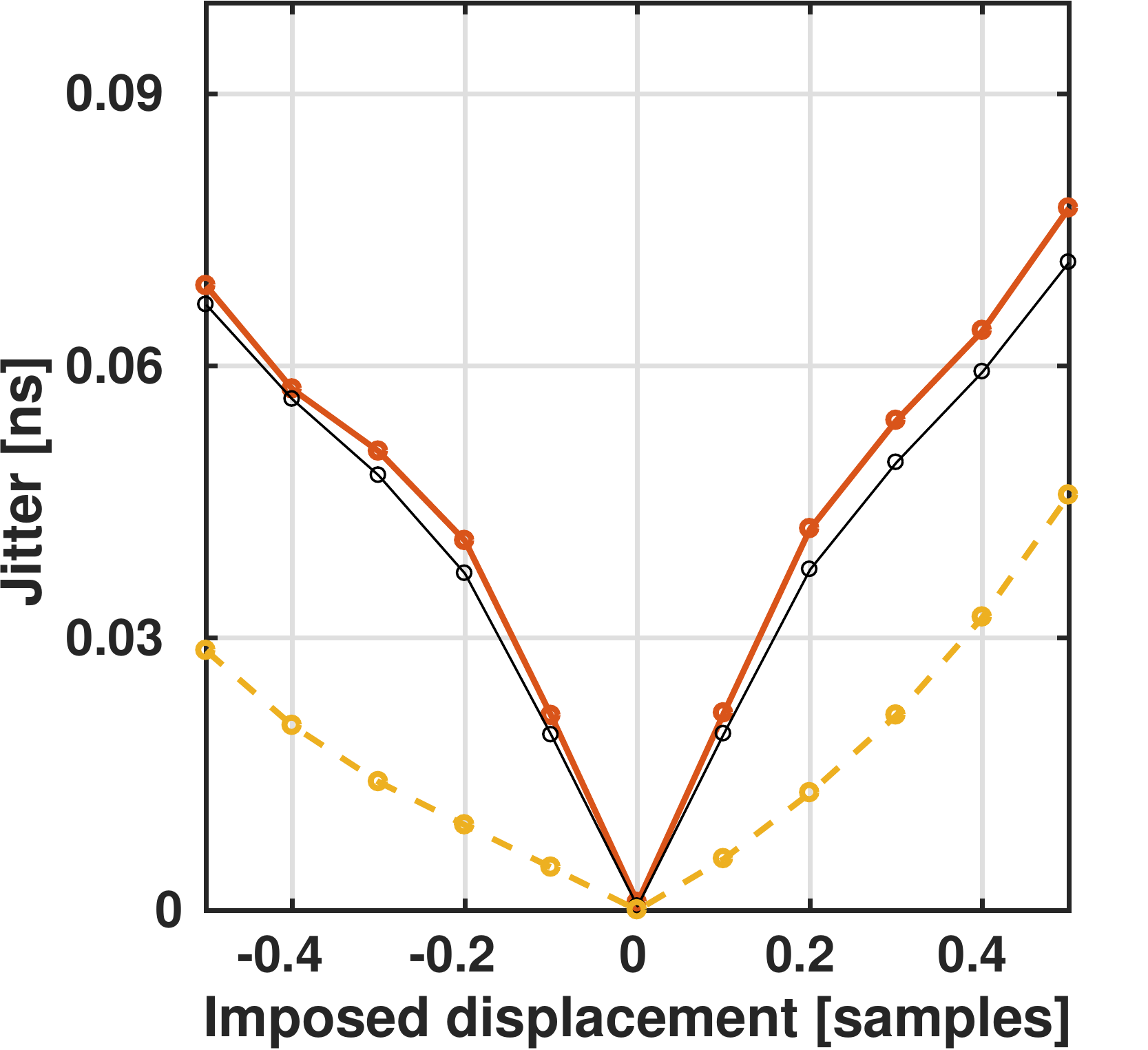}}
\put(0.4,0.073){\scriptsize (a)}
\put(0.91,0.07){\scriptsize (b)}
\end{picture}
\caption{a) Bias between the known and estimated displacement based on the 50 speckle realizations is shown as a function of the displacement value. Between displacements -0.5 to -0.3, the bias is away from the direction of displacement. Between -0.3 to 0.5, the bias is towards the direction of displacement. The average bias varies from $\lambda/1702$ to $\lambda/2162$ for kernel lengths ranging from $4-50\lambda$. b) Jitter is calculated as the standard deviation of bias, also shown as error bars in (a). The jitter increases with the magnitude of the imposed displacement, the average value ranging from 0.018 and 0.045 ns for kernel lengths ranging from $4-50\lambda$.}

\label{fig:jitter}
\end{figure}

Fig.~\ref{fig:jitter}(a) shows the bias between the known and estimated displacement, as a function of the displacement value. As expected, the bias for the reference (no displacement) case was zero. For most of the considered displacement ranges, between -0.3 to 0.5 there is a bias away from the origin, i.e. when displacement towards the transducer is imposed (negative displacement), the estimated bias is towards the transducer (negative bias) and vice versa. However, in a displacement range between -0.5 to -0.3 the bias is away from the origin, i.e. displacements towards the transducer (negative displacement) are biased away from the transducer (positive bias). The average bias for all displacements is also summarized in Table.~\ref{tab:bias}. At smaller kernel lengths of $10\lambda$ (3 mm) and $4\lambda$ (1.2 mm), the average bias was calculated as 0.0094 spatial pixel, equivalent to $\lambda/1702$. For a large kernel length of $50\lambda$ (15 mm), the average bias in estimates reduced to 0.0074 pixel ($\lambda/2162$). 

For the -0.5 pixel displacement case, the bias in displacement estimation was the highest. At this displacement, one of the pixels within each subresolution scatterer takes the value of the background medium. However, the error for the +0.5 pixel displacement is lower. This may be due to the axial asymmetry of the imaging point spread function. 

The jitter, calculated as the standard deviation of bias, is shown as error bars in Fig.~\ref{fig:jitter}(a)
and it is also represented directly in Fig.~\ref{fig:jitter}(b) for clarity. In general the jitter increases with the magnitude of the imposed displacement and it is smaller for larger kernel sizes. The average jitter, summarized in Table.~\ref{tab:bias}, varies between 0.018 and 0.045 ns.

\begin{table}[ht]
\centering
\caption{Summary of error in subresolution displacement estimates}
    \begin{tabular}{ |c|c|c|c| }
        \hline
        \textbf{Kernel length} & \textbf{$4\lambda$} & \textbf{$10\lambda$} & \textbf{$50\lambda$} \\
        \hline
        \textbf{Bias (spatial pixels)} & 0.0094 & 0.0094 & 0.0074 
        \\
        \hline
        \textbf{Bias ($\lambda$)} & 1/1702 & 1/1702 & 1/2162 
        \\
        \hline
        \textbf{Jitter (ns)} & 0.045 & 0.042 & 0.018 
        \\
        \hline

    \end{tabular}
    \label{tab:bias}
\end{table}

\subsection{Subresolution shear shock wave displacements}

To model shear shock wave motion a total of 1160 tissue and scatterer maps were generated. Each realization corresponds to a snapshot in time of the moving tissue based on its pre-calculated position from the shear shock wave simulation. Then the Fullwave simulation tool was used to simulate ultrasound propagation within each of these maps and the RF data from each map was beamformed as described previously. Similar to what has been done previously in experiments~\cite{Espindola2017}, the displacement tracking was performed in an interframe manner between subsequent frames. This tracking approach increases the accuracy by minimizing decorrelation errors.
The displacements were estimated in the noiseless case and with 6 dB  white noise added to the beamformed RF data measured with respect to the local RMS amplitude. The results are summarized in Table.~\ref{tab:shock}. 

The tracked shear shock waveform (shown in red, Fig.~\ref{fig:shock}) is compared to the known displacements (shown in black, Fig.~\ref{fig:shock}) for the noiseless case in Fig.~\ref{fig:shock}(a) and the 6dB noise case in Fig.~\ref{fig:shock}(b). In both cases the measured displacement closely matches the imposed displacement. {The RMS error between the imposed and measured shear wave displacements was 0.036 pixel ($\lambda/446$) in the noiseless case and 0.096 pixel ($\lambda/166$) in the added noise case. Although not directly visible in Fig.~\ref{fig:shock}, the error in estimates tends to increase with the magnitude of the imposed displacement, which is consistent with the results in Fig.~\ref{fig:jitter}. Hence, the error in the estimates of the shock wave motion, where the interframe displacement ranges up to 5.2 pixels, is larger than the error shown in Fig.~\ref{fig:jitter}, where the displacement range is restricted to 0.5 pixels between frames.

 {A significant source of error is due to the constant speed assumption in delay-and-sum beamforming. A uniform speed of sound value is assumed for a brain map during the beamforming process, even though the speed is in fact heterogeneous. This is a widely studied problem in beamforming~\cite{anderson2000impact} and although this error is independent of the displacement model or the impedance flow method it is still reported as an error. Thus due to the constant speed of sound assumption in the beamformer, the  displacement in the brain map is underestimated or overestimated depending on whether the local speed of sound is higher or lower than the assumed uniform value of speed in the beamforming process.} To illustrate this effect the displacement estimate as a function of depth was calculated for two difference average sound speed values (1537 and 1552 m/s) for a uniform displacement of 5 pixels (Fig.~\ref{fig:c0error}). At the focal depth (31.2 mm), for example, the displacement error is 0.079 pixel ($\lambda/202$) for a 1537 m/s beamformer and 0.022 pixel ($\lambda/723$) for a 1552 m/s beamformer. Furthermore, this error varies significantly as a function of depth, in a way that is consistent with the arrival time error due to local sound speed variations, i.e. and increase in the estimates where the sound speed in the map is larger and a decrease in the estimates where the sound speed is lower. The error due to the sound speed discrepancy could be overcome by using a dynamically changing $c_0$ value in the beamforming, although this is not included in the present study. This demonstrates that the error in the numerical representation of displacement is smaller than the physical error introduced by the beamforming process.




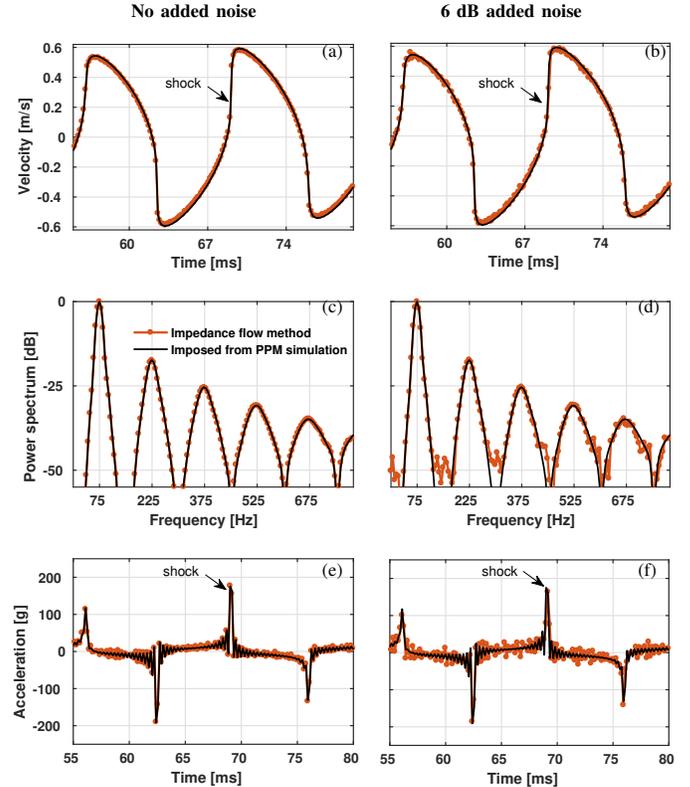
\begin{figure}[t]
\centering
\setlength{\unitlength}{0.475\textwidth}
\begin{picture}(0.9,1.2)(0,0.025)

\put(-0.05,0.02){\includegraphics[width=0.48\textwidth,trim=0 0 0 0,clip]{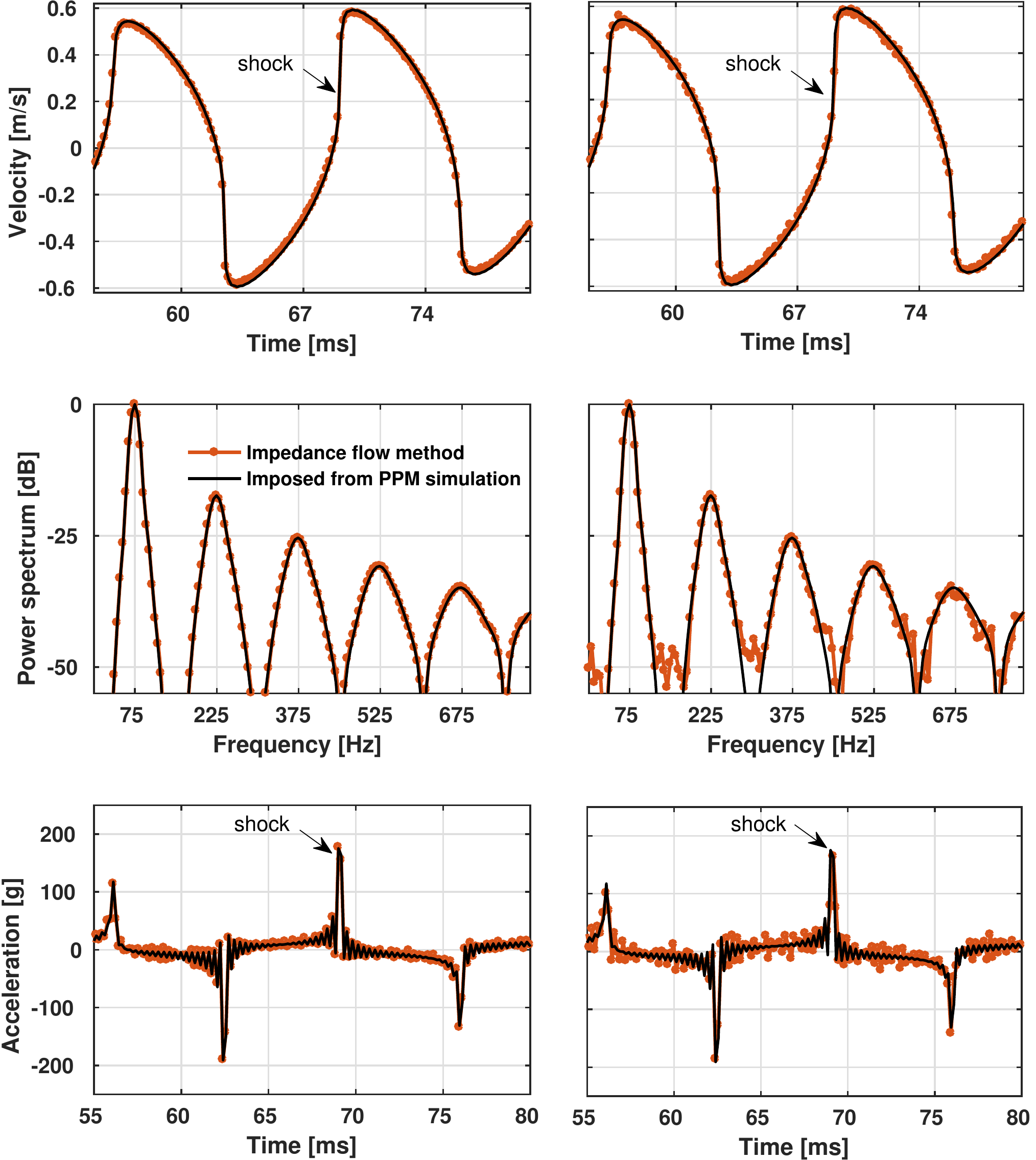}}

\put(0.13,1.19){\scriptsize \textbf {No added noise}}
\put(0.6,1.19){\scriptsize \textbf{6~dB added noise}}

\put(0.42,1.13){\scriptsize (a)}
\put(0.91,1.13){\scriptsize (b)}
\put(0.42,0.74){\scriptsize (c)}
\put(0.9,0.74){\scriptsize (d)}
\put(0.42,0.34){\scriptsize (e)}
\put(0.9,0.34){\scriptsize (f)}

\end{picture}

\caption{Interframe displacements are shown as the shear wave velocity signal along with the imposed shear wave for the a) no added noise and b) 6~dB added noise cases. 
The odd harmonic spectrum of the imposed cubic shock is retained both in the c) noiseless and d) added noise case. The shock acceleration is calculated using the interframe displacement estimates for the e) noiseless and f) added noise cases.}

\label{fig:shock}
\end{figure}

\begin{table}[ht]
\centering
\caption{Summary of error in shear shock wave displacement estimates}
    \begin{tabular}{ |c|c|c| }
        \hline 
        \textbf{Noise level} & \textbf{No added noise} & \textbf{6 dB} \\
        \hline
        \textbf{RMS error (spatial pixels)} & 0.036 & 0.096 \\
        \hline
        \textbf{RMS error ($\lambda$)} & 1/446 & 1/166  \\
        \hline
        \textbf{Error in peak acceleration ($\rm \%$)}  & 1.51 &  4.53
        \\
        \hline

    \end{tabular}
    \label{tab:shock}
\end{table}

\begin{figure}[h]
\setlength{\unitlength}{0.475\textwidth}
\begin{picture}(1,0.43)(0,0.15)
\centering

\put(0.08,0.38){\includegraphics[width=0.436\textwidth,trim=135 350 130 350,clip]{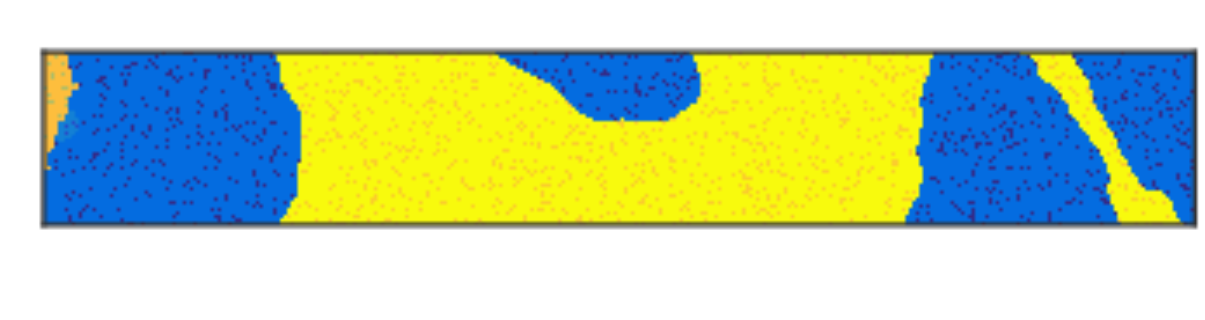}}
\put(0.044,0.117){\includegraphics[width=0.455\textwidth,trim=0 0 0 0,clip]{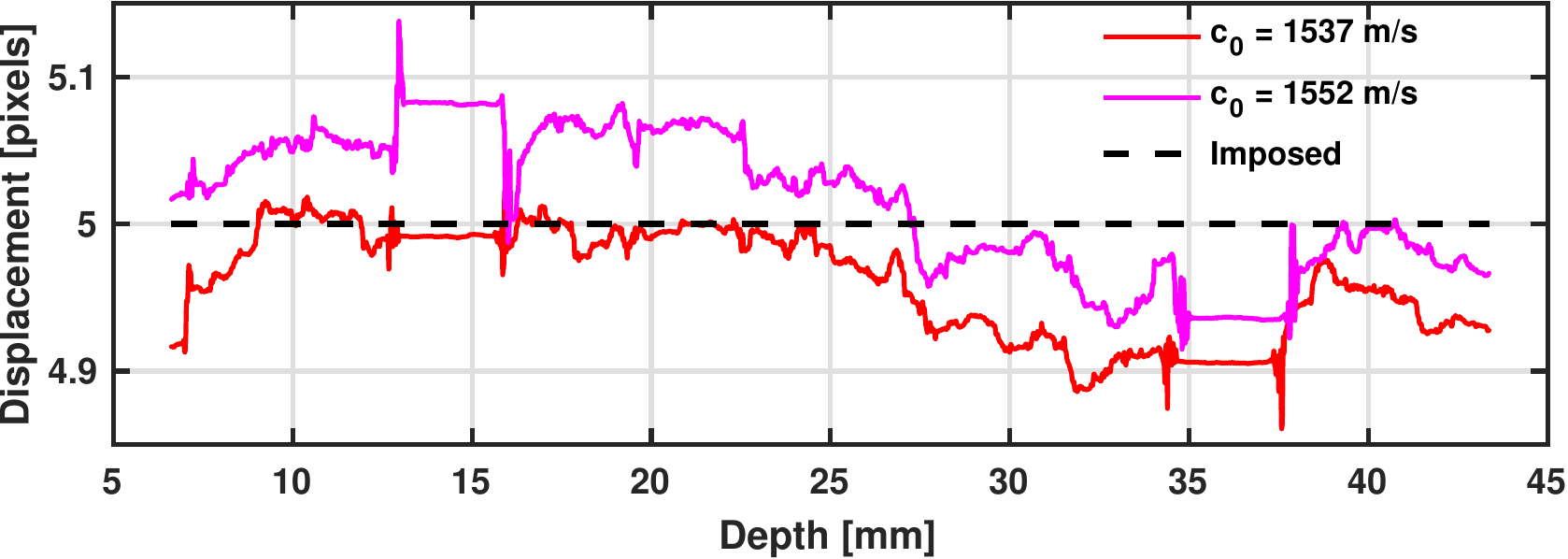}}

\end{picture}

\caption{Variation of estimates of the imposed 5 pixel displacement along the field depth for the heterogeneous brain map, using the average field sound speed of 1537 m/s and the local speed of 1552 m/s in the focal region.}

\label{fig:c0error}
\end{figure}

As can be seen from the accompanying power spectra in Figs.~\ref{fig:shock}(c,d), the characteristic odd harmonic spectrum arising from the cubic nonlinearity of the elastic deformation is clearly observable at least until the ninth harmonic in both cases.

The acceleration waveform calculated  using a Fourier spectrum method is shown in Figs.~\ref{fig:shock}(e,f). 
Acceleration was calculated from the tracked interframe displacements using the Fourier spectra as ${a(t)=\mathcal{F}^{-1}(2{\pi}i f \mathcal{F}(v(t))})$, where $a(t)$ is the acceleration time signal, ${i=\sqrt{-1}}$, $f$ is spectral frequency and $v(t)$ is the velocity time signal. The error in peak acceleration was measured to be 1.51\% in the noiseless case and 4.53\% in the 6 dB noise case. 


\subsection{ {Experimentally measured shear shock wave displacements from the \textit{in situ} brain}}

 {Experimentally measured shear shock waves in the \textit{in situ} brain were imposed as subresolution displacements in the acoustical brain maps. This process was similar to the imposition of PPM-simulated shear shock waveform (Fig.~\ref{fig:shock}) in the previous section except that experimental estimates of displacements were used. A craniectomy was performed on a freshly extracted porcine head. Subsequently, a plate source was attached which was oscillated vertically using a 40~Hz sinusoidal pulse, to propagate shear waves in the brain. Images of the brain moving under the propagating shear waves were collected using a custom ultrasound sequence at 2200 images/s.}



 {The fully developed shear shock wave observed inside the brain at a propagation distance of 6.7 mm from the source (shown in black, Fig.~\ref{fig:experimental}(a)), which was chosen to impose as displacement in the acoustical brain maps. The experimental velocity waveform consists of 570 samples each of which were converted to interframe displacements and imposed on the brain map shown in Fig.~\ref{fig:brain}(c), to generate 570 displaced acoustical maps. Fullwave simulations were performed on each of the displaced fields to generate the numerical RF data at the mid-lateral position. The simulated RF data from all the 570 displaced brain fields were then tracked using the adaptive tracking algorithm as described previously in II,E. The tracked displacements are shown in terms of particle velocity in Fig.~\ref{fig:experimental}(a) (red line, labeled 'impedance flow method') and is seen to closely match the imposed experimental shear shock waveform. In the accompanying power spectra (Fig.~\ref{fig:shock}(b)), the odd harmonic content is captured up to at least the ninth harmonic at -30 dB. Moreover, the peak acceleration of 110$g$ at the shock front at 142 ms is captured with an error of $<$ 0.843\%. }

\subsection{ {Discussion on nonlinear shear wave modeling}}

 {The recent discovery that shear shock waves can form in soft solids~\cite{catheline2003observation} and the brain~\cite{Espindola2017} has generated interest in modeling and simulating this practically unstudied biomechanical behavior. In general, shear wave deformations in soft solids have been studied using the stress-strain response based on a strain-energy-density-function \cite{Ogden1997}.
Based on Landau's model \cite{Landau1986} which consists of third- and fourth- order terms of Lagrangian strain tensor invariant, a nonlinear shear wave equation was derived \cite{zabolotskaya2004modeling} for linearly-polarized waves. This nonlinear model, originally developed for lossless media, was subsequently extended to model nonlinear shear wave propagation in attenuating and dispersing soft solids \cite{Tripathi2019_PPM1D,Tripathi2019_PPM2D}. The model is capable of imposing empirical attenuation power laws while preserving the Kramers-Kronig causal dispersion relations. A more generalized 3D model beyond the linearly polarized restriction is an ongoing work.} 

 {Although the field of nonlinear elastodynamics is vast, there are few reports of numerical modeling of shear shock waves. Finite element methods are generally designed to model nonlinear elastic behavior \cite{Zienkiewicz2005,Ye2017b}  however, there are no reports of finite element methods successfully modeling shear shocks. Thus for transient waves in general, and especially shock waves, methods that are designed to be flux conservative, such as finite volume methods seem to be more appropriate \cite{Leveque2002-a}. Since the flux conservative approach guarantees the correct shock speed, we have chosen a piecewise-parabolic finite volume method to numerically resolve our nonlinear shear wave model \cite{Tripathi2019_PPM1D,Tripathi2019_PPM2D}. }
 {In this paper, we have not used the most recent conventional time models \cite{Tripathi2019_PPM1D} instead opting to use a previous nonlinear model which uses a retarded time frame formulation (i.e., spatial frame of reference moves with the wave form) \cite{Tripathi2017}. The older model has been extensively validated experimentally and it includes a parameter estimation study that was performed to determine the nonlinear viscoelastic properties in fresh brain \cite{Espindola2017}. Ongoing studies will  determine the viscoelastic properties for the conventional time formulation, at which point the new model can easily replace the old one. Nevertheless, since the shear shock waveforms in the new and old models have essentially the same shape and the same spectrum and the effect on displacement estimation is likely to be insignificant.}

\subsection{Computational time for modeling equivalent displacements on a refined grid}

 {The simulations presented here had a run time of approximately 50 min on a single core of an Intel Xeon E5-2680 v3 2.50-GHz CPU. Without the proposed subresolution displacement method, in order to obtain a resolution equivalent to 0.0094 spatial samples (the average error in the displacement model from Fig.~\ref{fig:jitter}(a)), we would need to refine the spatial grid by a factor of 106. This would increase the computation time by a factor of \num{1.2e6}.
}


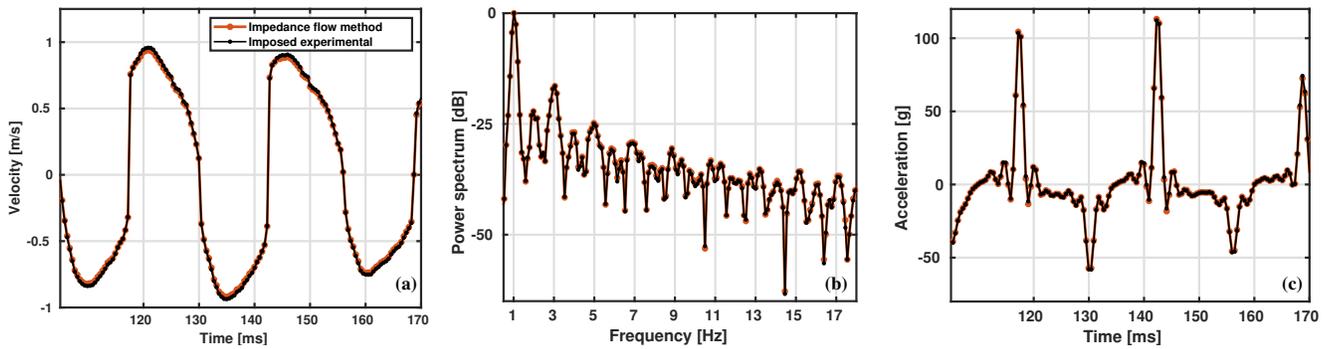
\begin{figure*}[h]
\centering
\setlength{\unitlength}{\textwidth}
\begin{picture}(1,0.25)(0,-0.0)

\put(0.03,0.0){\includegraphics[height=4.5cm,trim=0 0 0 0,clip]{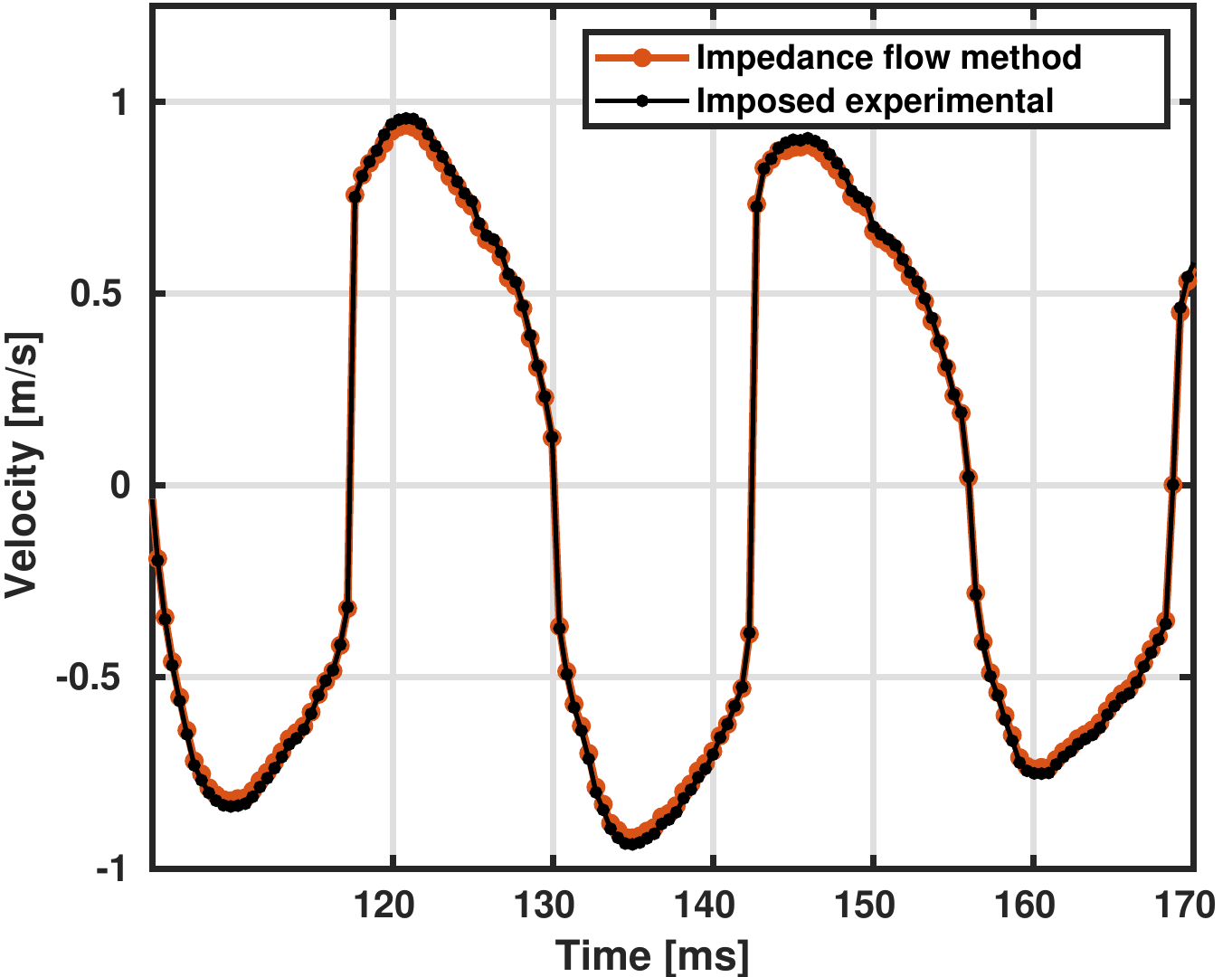}}
\put(0.35,0.){\includegraphics[height=4.5cm,trim=0 0 0 0,clip]{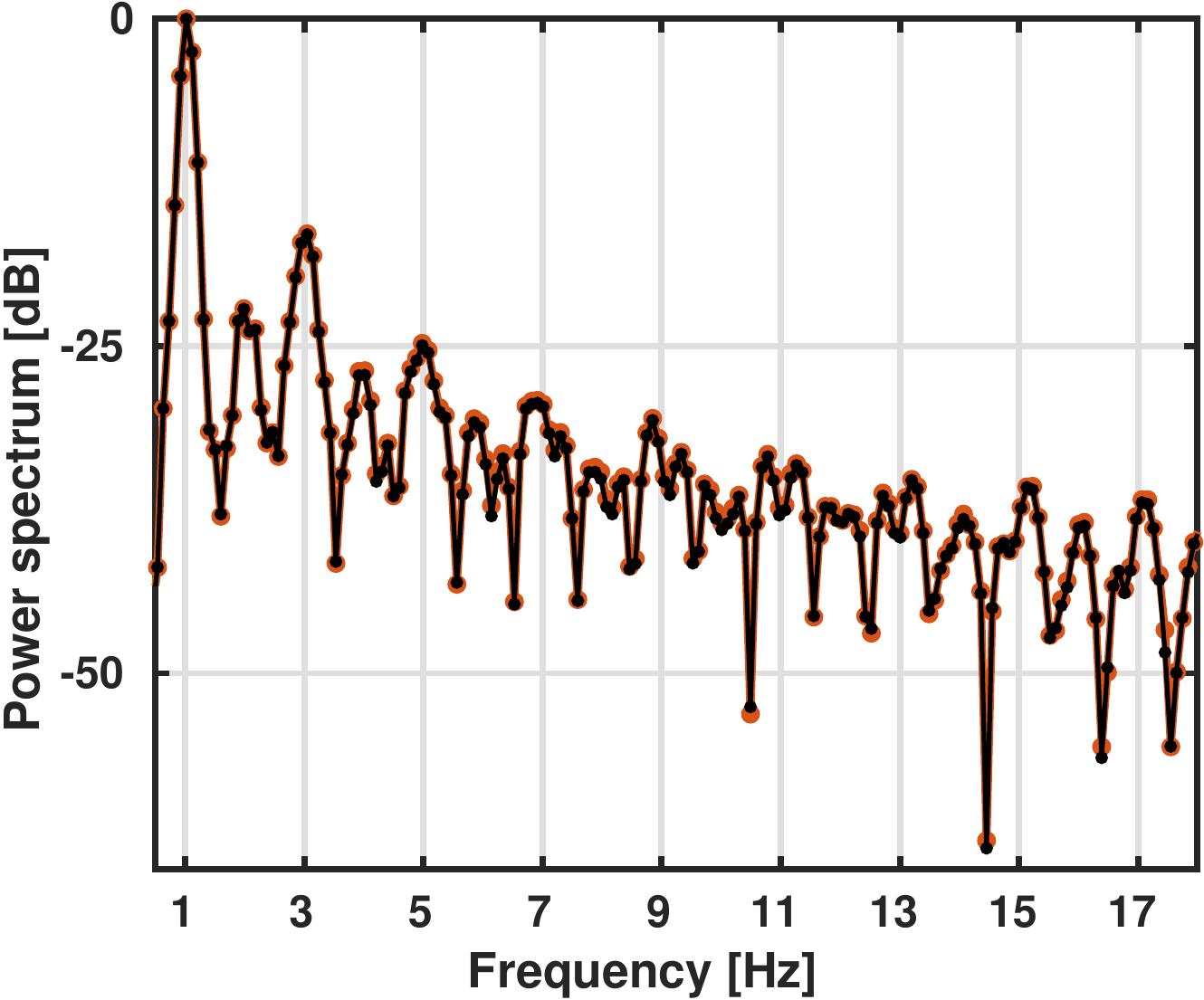}}
\put(0.67,0.){\includegraphics[height=4.5cm,trim=0 0 0 0,clip]{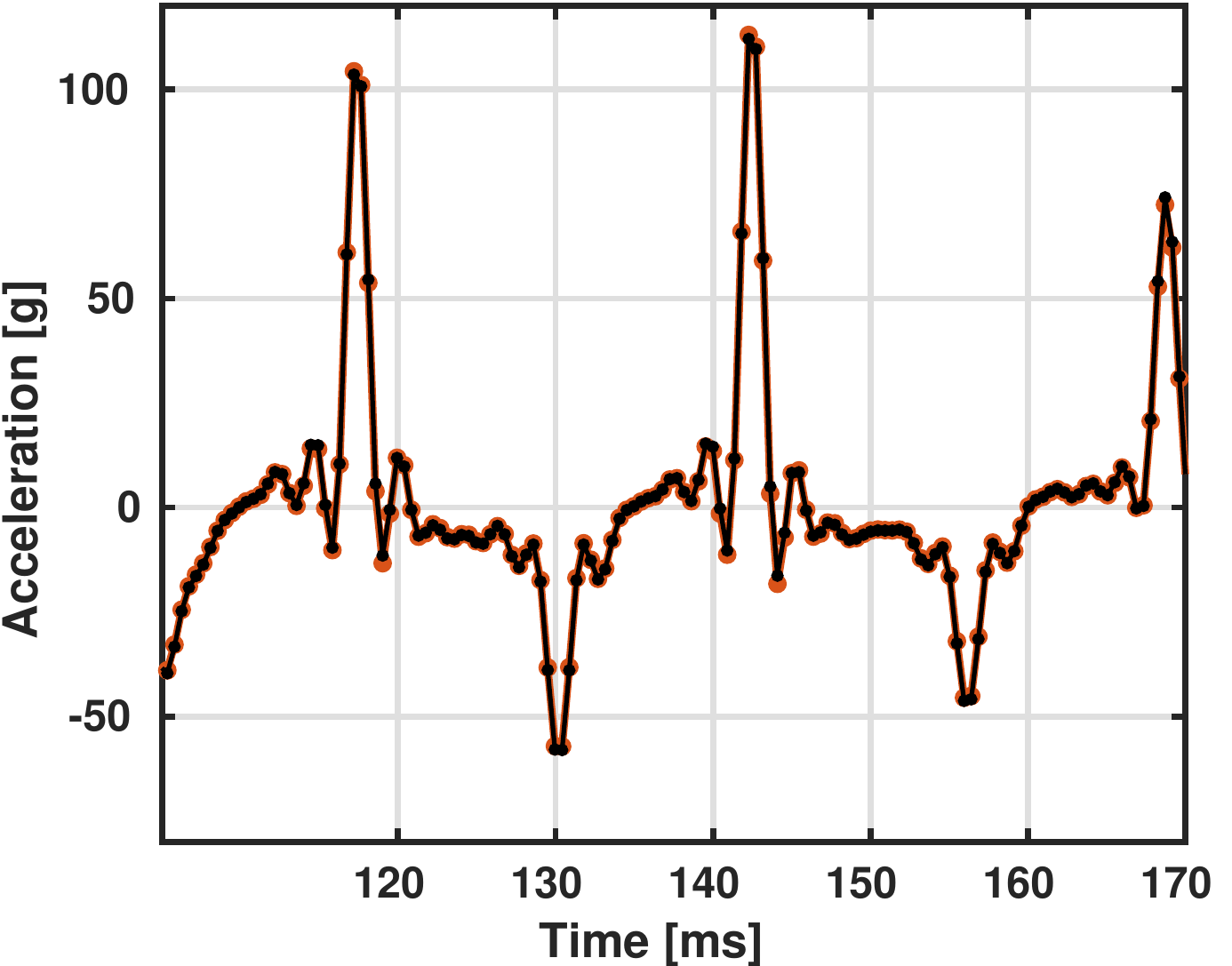}}

\put(0.31,0.04){\scriptsize \textbf{(a)}}
\put(0.62,0.04){\scriptsize \textbf{(b)}}
\put(0.95,0.04){\scriptsize \textbf{(c)}}

\end{picture}

\caption{a) Experimentally measured shear shock wave displacements were imposed in the brain maps as subresolution displacements using the proposed impedance flow method. The tracked velocity waveform based on Fullwave simulated RF data is in excellent agreement with the imposed waveform. b) The odd harmonic spectrum of the experimental shock wave is retrieved in the waveform tracked from brain maps. c) The shock front acceleration peak was retained in the tracked waveform with an error of $<$ 0.843\%.}

\label{fig:experimental}
\end{figure*}



\section{Summary and conclusion}


A previously described impedance flow method~\cite{pinton2017continuous} was generalized to represent subresolution motion on finite grids and so that the motion of the brain during shear shock wave propagation can be described within ultrasound imaging simulations. The generalization enables implementing displacement in an arbitrary distribution of impedance values, without the restriction of assuming a two-pixel scatterer model. 

 {This subresolution displacement method can be seen as an analogy of anti-aliasing in computer graphics \cite{korein1983temporal,crow1981comparison,damera2009display,fuchs1986fast}. In the example shown in Fig.~\ref{fig:scat}(a-b), the impedance flow operation performed to impose the 0.2 pixel displacement is effectively the same as imposing a 1 pixel displacement on a grid that is refined by a factor of 5 and subsequently downsampling the impedance map back to its original grid size. The downsampling operation essentially assigns to each pixel, the average intensity of its sub-pixels on the intermediate refined grid. }

 {Note that impedance flow can be implemented by changing either or both of density and speed of sound. For the sake of simplicity, only the sound speed has been varied, while fixing a constant density for the medium. Generally speaking, sub-pixel flow would be implemented for all of the physical parameters of the field including nonlinearity and attenuation. The medium was assumed to be  linear and non-attenuating to avoid potential confounding effects}


It was shown here that the error in representing displacements measured by a 5.2 MHz imaging transducer was  $\lambda/1702$, which is less than 0.2 $\mu$m. This accuracy was more than sufficient to represent shear shock wave propagation in the brain with respect to key characteristics such as the third harmonic amplitude of the shock wave and the acceleration at the shock front. The method was also shown to accurately represent the cubic nonlinearity at the steep shock front of the imposed wave beyond the ninth harmonic, which is the limit of what is currently achievable experimentally. The bias in the displacement estimate was seen to increase with the displacement, which is to be expected. However, the bias was shown to be sensitive to the speed of sound value used in the beamformer, independent of the subresolution displacement model. By using the local sound speed value in the region of interest for displacement tracking, the bias in the displacement estimates was shown to decrease. Although not re-implemented here, the displacement accuracy can be further improved by including a calibration step as described previously \cite{pinton2017continuous}.  {Although the error characterization here was performed using 2D Fullwave simulations, we expect that the wave physics describing relative phase shift i.e., displacements will remain unchanged in a 3D imaging scenario. There may be however a need to re-calibrate the scatterer brightness differently in a 3D field, so that it has the same values as soft tissue. }

 {Additionally, it was shown that experimentally measured shear shock wave displacements in the \textit{in situ} brain could be accurately represented in the acoustical maps in terms of the shock front characteristics.}
 {The subresolution displacement method along with Fullwave simulations is a versatile platform to test and validate transmit beam sequences customized for measuring shear shock wave displacements and to validate motion tracking algorithms for tracking in different parts of the body. For instance, the Fullwave simulations can include the skull within the imaging medium and can also include a spatially varying displacement field that mimics experimentally challenging scenarios. This method provides the capability to simulate realistic 2D or 3D shear shock wave displacements in soft tissue, which can be used to rapidly design and evaluate imaging sequences and tracking algorithms for shock wave detection. Such capabilities are especially useful in the context of TBI studies where the scope for experiments is restricted.} 



 {Although the impedance flow method is demonstrated for discretization of acoustic impedance in an ultrasound imaging field, the concept is broadly applicable to model sub-pixel flow of any physical quantity on the finite difference grid.} Furthermore this approach could be extended to represent motion in 2D and 3D, by ``flowing'' the impedance along the desired displacement direction. 
The proposed method can be used generally to model different types of motion or deformation including tissue motion due linear shear waves in elastography or physiological motion with applications to motion filtering. This method can be applied to the larger body of work in finite difference simulations so that small displacements can be modeled in the context of ultrasound imaging, phase aberration, reverberation, focused ultrasound, multiple scattering in bubbly media, coherence imaging, etc~\cite{pinton2011effects,pinton2011erratum,joshi2017iterative,pinton2014spatial}. Besides ultrasound imaging, the technique can be applied to applications involving small displacement of any physical quantity such as stress, strain, temperature or electric potential.

\end{document}